\documentclass[10pt,letterpaper]{article}
\usepackage[top=0.85in,left=2.75in,footskip=0.75in]{geometry}

\usepackage{changepage}

\usepackage[utf8]{inputenc}

\usepackage{textcomp,marvosym}

\usepackage{fixltx2e}

\usepackage{amsmath,amssymb}

\usepackage{cite}

\usepackage{nameref,hyperref}

\usepackage[right]{lineno}

\usepackage{microtype}
\DisableLigatures[f]{encoding = *, family = * }

\usepackage{rotating}

\usepackage{subfigure}

\usepackage{xcolor}

\usepackage{parskip}

\raggedright
\usepackage[aboveskip=1pt,labelfont=bf,labelsep=period,justification=raggedright,singlelinecheck=off]{caption}

\makeatletter
\renewcommand{\@biblabel}[1]{\quad#1.}
\makeatother

\date{}

\usepackage{lastpage,fancyhdr,graphicx}
\usepackage{epstopdf}
\fancyheadoffset[L]{2.25in}
\fancyfootoffset[L]{2.25in}
\begin{document}
\vspace*{0.35in}

\begin{flushleft}
{\Large
\textbf\newline{A numerical study of the the  response  of  transient  inhomogeneous  flames  to pressure  fluctuations  and negative stretch in contracting hydrogen/air flames}
}
\newline
\\
Nadeem A. Malik$^1$, T. Korakianitis$^2$, and T. L$\o$v\aa s$^3$, T. 
\\
\bigskip
$^1$Department of Mathematics and Statistics, King Fahd University of Petroleum and Minerals,
P.O. Box 5046, Dhahran 31261, Saudi Arabia\\
$^2$St. Louis University, Parks College of Engineering, Aviation and Technology, 
3450 Lindell Boulevard, St. Louis, Missouri 63103, USA.\\
$^3$Norwegian University of Science and Technology, 
Dept. of Energy and Process Technology, Kolbj\o rn vei 1b, 
NO-7491 Trondheim, Norway\\
\bigskip

{\em \small Corresponding author E-mail: namalik@kfupm.edu.sa, and nadeem\_malik@cantab.net}

\end{flushleft}

\setlength{\parskip}{5pt}
\setlength\parindent{0pt}

\section*{Abstract}
Transient  premixed  hydrogen/air  flames  contracting
through inhomogeneous  fuel distributions  and  subjected to
stretch and pressure oscillations are investigated numerically   using
an implicit  method which couples  the fully compressible flow  to the
realistic chemistry and  multicomponent transport properties. The  
impact of increasing {\em negative} stretch  is  investigated  through  
the   use  of  planar, cylindrical and  spherical geometries,   and  a 
comparison  with the results from {\em positively} stretched expanding 
H2/air flames (\cite{MALIK2010}) and CH4/air flames (\cite{MALIK2012a}) 
is made. A flame relaxation number $n_R=\tau_R/\tau_L$ ($\tau_R$ is the 
time that the flame takes to return to the mean equilibrium conditions 
after initial disturbance; $\tau_L$ is a flame time scale) decreases 
by 10\% with  increasing  {\em negative}  stretch,  in  contrast to  
the two expanding flames where  $n_R$ decreased by 40\% with  increasing 
{\em positive} stretch. $n_R$ appears to much more sensitive to 
variations in positive/negative curvature than to the 
thermo-chemistry of different flame types. $n_R$ may thus be a useful
indicator of the strength of flame-curvature coupling.  
The spectra of pressure fluctuations 
$E_p(\omega)$ scale  close  to  $\sim\omega^{-3}$,  which  is steeper  
than in the expanding H2/air flames where  $E_p(\omega)\sim\omega^{-2}$. 
Rapid transport ('flapping')  
of the  flame front  by the  largest convective velocities induced by 
the random pressure fluctuations is prominent  because of the lack of 
gas velocity ahead of the flame, $u_g=0$. 'Memory effects' between fuel
consumption and the rate of  heat release is obscured by the flapping,
although the contracting flames display short  time-lags in the mean. 
A spectrum  of  species thickness  $\{l_k\}$  exists  
such that  species formed in thin reaction layers are generally not 
disturbed except with respect to their peak concentration, and 
species formed in thick reaction layer are disturbed along their
entire length.

\bigskip
\noindent PACS numbers:  47.27.E?, 47.27.Gs, 47.27.jv, 47.27.Ak, 47.27.tb, 47.27.eb, 47.11.-j 


{\bf keywords:} flame;premixed; pressure fluctuations; stretch; flame velocity; 
chemical kinetics; curvature;multicomponent transport; methane; 
direct simulations; implicit.



\section{Introduction}
Most naturally occurring flames are turbulent and unsteady and occur in 
the  presence  of significant  local stretch. Furthermore, 
The response  of the flame  kernel  to pressure  fluctuations  can lead  
to violent amplification of the heat release, while lean gas mixtures 
can be difficult to  ignite and sustain, with the possible consequence 
of self-induced oscillations. The coupling between the
flame structure and the unsteady flow field is therefore of 
fundamental importance both  to the theory of flames and to practical 
combustion systems.

But direct measurements of the local 
flame structure and its response to stretch has proved difficult. As a 
consequence cylindrically and spherically symmetrical flames have often 
been studied because these geometries provide simplifications from which   
information  regarding  fundamental aspects of flames can still be 
extracted.

Outwardly  propagating  flames  occur  widely such  as  in  automotive
engines where  fuel/air mixtures are  ignited inside a  cylinder under
high pressure  to form a  small approximately spherical  expanding
flame  that releases heat to produce useful work, as seen in the
images inside an engine captured by \cite{HEYWOOD1988}. 
Expanding  flames have been studied  more often 
than  implosions  as  they   are  easier  to  produce  experimentally;
They  have  been studied  experimentally by~\cite{DOWDY1990,
TAYLOR1991, KWON1992, TSENG1993, KARPOV1997, KELLEY2009}  and
numerically by~\cite{BRADLEY1996, MALIK2010} and~\cite{MALIK2012b}.

Inwardly propagating flames are important because they are not subject
to strain  due to the zero gas velocity ahead of the flame, and stretch is
therefore due to  the curvature alone and so  the effects of curvature
can be studied in isolation.   This is important because curvature can
have  a dominant effect  in many  combustion phenomena  as well  as in
other    powerplant    flows, see for example    
\cite{TK-pcd-1993,    TK-curveff-1993,
TK-inlet-2007, Pilidis2006a, TK-Idres-AE-2010}. Experimentally, 
implosions are difficult to produce and little data is
available;  exceptions  though  are  the  experiments  of
al~\cite{BAILLOT2002}  and~\cite{GROOT2002}.   Imploding
methane/air flames  have been  investigated numerically by
~\cite{BRADLEY1996}, ~\cite{KELLEY2009} and~\cite{MALIK2012b}.

The last twenty years have seen great progress in the understanding of
flame  interactions  with physical  oscillations  of different  types.
Quite often models are adopted with approximations in the modeling of
the   flow    and/or   the   chemistry,    e.g.    1-step   chemistry,
\cite{TEERLING2005};  parametric  variations   in  only  one  physical
variable    such   as    the    equivalence   ratio
\cite{MARZOUK2000,
SANKARAN2002},   induced   pressure  fluctuations, \cite{TEERLING2005},
harmonic velocity fluctuation, \cite{LIEUWEN2005}, and  flame response
to    unsteady     strain    and    curvature     have    also    been
explored, \cite{NAJM1997}.  The impact  of pressure  fluctuations upon
combustor  stability  have been  highlighted in  a  number of  studies
e.g.~\cite{CANDEL2002,LUFF2006},  and  there   is  evidence  that  the
intrinsic thermochemical structure may lead to pulsating instabilities,
\cite{CHRISTIANSEN2002, GUMMALLA2000, LAUVERGNE2000}.
Large Eddy  Simulations with  a flame surface  density model  has been
used to investigate expanding flames, \cite{IBRAHIM2009}.

Most  of  the studies  noted  above  focus  on particular  effects  in
isolation, but there is also a need to examine more complex situations
where many unsteady processes are occurring at the same time, which is
closer  to  reality. Therefore,  in this study we investigate the
{\em  combined} effect  of simultaneous  fluctuations in  pressure and
fuel concentration in unsteady  finite thickness flames.

Although major advances in  computer simulations of combustion systems
have  been made in  recent years,  detailed computational  studies are
still   comparatively  few,   e.g. \cite{BRADLEY1996,  LINDSTEDT1998b,
MEYER2001,  TK-emissions-2004,  TK-Dyer-2007,  MALIK2010, MALIK2012b}.  
With  the
need to interrogate complex reacting  systems in greater detail, it is
recognised that  a step  forward in computational  combustion 
requires  the inclusion  of more  realistic chemistry  coupled  to the
compressible flow, as highlighted in~\cite{LU2009, MALIK2010, 
MALIK2012a, MALIK2012b}.

The  need  for the  inclusion  of  realistic  chemistry is  especially
important  in view  of the  need to  predict emissions  of green-house
gasses and particulates  from industrial devices with a  fair degree of
accuracy and reliability, as  noted by~\cite{LOVAS2010}.
New challenges are related to the yet unknown behaviour of alternative
fuels, such as bio-fuels.   Although overall green-house gas emissions
are low for  biomass fuels, these fuels emit  higher concentrations of
heavy  metals and  particulates which  can be  a hazard  to industrial
devices,  such as  in  automotive  engines where  the  role played  by
differential  diffusion  in  species  and  particle  transport  is  of
special concern, \cite{LOVAS2010,MAUSS2006,MARCHAL2009}.

The recent progress in numerical software and in computer technology 
now offers the possibility to couple the detailed chemistry to the 
compressible flow.
Large  Eddy Simulations  are  the  methods of  choice  in large  scale
numerical simulations in practical applications because they provide an
acceptable compromise between grid resolution and the overall size of 
the system, e.g.~\cite{MALALA2008}   investigate
recirculation and  vortex breakdown of swirling flames  using LES, and
Luo et al have used  LES  to  investigate  interactions between  a
reacting plume  and water spray, \cite{XIA2008}, and the aeroacoustics
of     an    unsteady     three-dimensional     compressible    cavity
flow, \cite{LAI2007}.

At  a  more  fundamental  level, Direct  Numerical  Simulations  using
explicit  methods have  matured  in  the last  fifteen  years and  are
providing useful data, \cite{THEVENIN2002,CHAKRABORTY2004,JIANG2007, 
LIGNELL2008, SCHROLL2009,CHEN2006,GOU2010}. Although they have had 
stability problems,
significant progress has been made recently in chemical stiffness
removal and multi-dimensional scale integration in the studies
mentioned above. Nevertheless, it is common practice to use reduced
chemical schemes in these codes in order to improve stability.

Implicit methods  have been receiving some attention  recently as they
are  well known  to  provide  stable solvers  even  for stiff  systems
although  at  the  cost  of  large memory  requirements.   Smooke  and
co-workers  have  developed implicit  compact  scheme  solvers for  2D
non-reacting flows,  \cite{NOSKOV2007}, and for 1D  reacting flows with
simplified  global reaction  source term, \cite{NOSKOV2005}.
Malik  has developed an implicit finite volume method, 
TARDIS (Transient Advection Reaction Diffusion Implicit
Simulations), \cite{MALIK2010,MALIK2012a,MALIK2012b}, 
featuring  the coupling of the compressible  flow to the
comprehensive chemistry and the detailed multicomponent 
transport properties of the mixture, thus resolving all the convective 
and chemical length and  time  scales which  is essential for studying  
stiff and unsteady reacting systems.  

TARDIS thus  provides a platform 
for studying in detail regimes of combustion which were previously 
difficult to explore due to  the limitations of experimental and 
other numerical methods; for example, unsteady combustion processes
in automobile engines have not been explored numerically or 
experimentally with full chemistry. One of the main purposes of this work
is to deomstrate the capability of TARDIS to such complex systems.

An area of special interest in  this study is the regime of relatively
high Karlovich number $K=\kappa\tau_L \geq 1$, where 
$\kappa=A^{-1}dA/dt$  is the stretch, the rate of change of an 
elemental area $A(r)$  surrounding 
a point on the surface of  the  flame, and  $\tau_L$ is  a flame  time
scale. In this limit, the convective fluctuations can penetrate inside
the internal structure of the flame thus disturbing the flame structure 
and the system  is then essentially unsteady. Little is known  of  the  
flame  response  in  this limit  and yet  its occurrence is widespread 
because most  combustion occurs under turbulent conditions, such as in 
internal combustion engines, when the smaller convective scales are 
able to penetrate  inside the flame
kernel.   Many  models  of  turbulent  flames often  assume  that the
turbulent flame front consists of a collection of laminar flamelets so
that  the  steady  laminar  flame structure  is  assumed  to remain 
undisturbed. But clearly such an assumption cannot remain valid 
when the convective scales penetrate inside the flame kernel.

The  aim of  this work  is two-fold.   First, we  use TARDIS  to study
stoichiometric   hydrogen/air   {\em   inwardly}   propagating   flames
(implosions)   in  cylindrical  and   spherical  geometries   at  mean
atmospheric  pressure -- essentially small fuel pockets being consumed. 
We  explore the  coupling of  the thermochemical
flame   structure  to   pressure  oscillations   in  the   context  of
inhomogeneous fuel distributions and  for different levels of negative
stretch (flow divergence), and how this affects properties of the flame  
such as the flame  speed and  flame  structure.  

Second,  the results  will  be compared  to  the results  from
\cite{MALIK2010, MALIK2012a} who carried out related studies  
on {\em expanding}  hydrogen/air and methane/air flames.

\section{Inwardly propagating flames}

Inwardly propagating flames have not been studied as much as expanding
flames because they are difficult to produce experimentally. Exceptions 
though are~\cite{BAILLOT2002}  and~\cite{GROOT2002}; 
Baillot et al  produced  small spherical bubbles of methane-air about 
1~mm  diameter (which have high curvature).  Numerical  simulations of
imploding  flames are  also scarce;  \cite{BRADLEY1996}
investigated  the   methane/air  system  using   a  simplified  4-step
mechanism  and a  simplified set  of  balance equations  in which  the
momentum  equation was  neglected under  the assumption  that pressure
gradients  were negligible;  and \cite{MALIK2012b} has 
investigated stretched flame velocities in  different geometries for
different fuels and modes of propagation.

Inwardly propagating laminar  flames are not subject to strain
due to the zero velocity ahead  of the flame. Stretch is therefore due
to the curvature alone and so  the effects of curvature can be studied
in  isolation.

A flame  front velocity $u_f$ is  defined as the  physical movement of
the flame  in space, and is given  by the sum of  the burning velocity
$s_n$  and the  gas velocity  $u_g$,  $u_f=s_n+u_g$.  In  the case  of
inwardly propagating  flames where $u_g=0$  ahead of the flame  in the
fresh gas mixture, we have $u_f  = s_n$. In the current study, there
is a third  component $u'_p$ induced by the  pressure fluctuations, thus
$u_f =  s_n+u'_p$ ahead  of the  flame. Note that  the pressure 
fluctuations exist throughout the domain and therefore $u'_p$ can  be 
non-zero anywhere in the domain, even ahead of the flame where the
velocity is usually expected to be zero.

\section{Method}

The     current     study     features     an     implicit     method,
TARDIS (see \cite{MALIK2010,MALIK2012a,MALIK2012b}), 
in  which  the  balance equations  of  mass,
momentum,  energy  and  chemical  species,  together  with  the  state
equation  for ideal  gas, are  solved in  an implicit  framework  on a
staggered  grid  arrangement.   The  governing balance  equations  are
functions  of  time, $t$,  and  one  spatial  coordinate, $x$  or  $r$
(radius), only.  The governing equations are,
\begin{eqnarray}
 \displaystyle{{\partial\rho}\over        {\partial       t}}       &+
 \displaystyle{1\over         r^\alpha}{{\partial         r^\alpha\rho
 u}\over{\partial  r}}   &=  0\\  \displaystyle{{\partial\rho  u}\over
 {\partial  t}} &+  \displaystyle{1\over  r^\alpha}{{\partial r^\alpha
 \rho     u^2}\over{\partial      r}}     &=     -\displaystyle{1\over
 r^\alpha}{{\partial     r^\alpha     J_\nu}\over\partial     r}     -
 \displaystyle{{\partial   P}\over  \partial   r}  +   \chi^\alpha  \\
 \displaystyle{{\partial\rho     e}\over      {\partial     t}}     &+
 \displaystyle{1\over         r^\alpha}{{\partial         r^\alpha\rho
 ue}\over{\partial  r}}  &= -\displaystyle{1\over  r^\alpha}{{\partial
 r^\alpha  J_e}\over\partial   r}  \nonumber  \\   &&  -  \sum_{k=1}^N
 \displaystyle{{1\over  r^\alpha}{{\partial}\over \partial r}[r^\alpha
 h_k(-J_k+J_n)]}        \nonumber\\         &&        +        {P\over
 r^\alpha}\displaystyle{{\partial  r^\alpha  u}\over  {\partial r}}  +
 \Gamma^\alpha   +   \mu\Psi^\alpha   \\   \displaystyle{{\partial\rho
 Y_k}\over  {\partial t}} &+  \displaystyle{1\over r^\alpha}{{\partial
 r^\alpha  \rho  uY_k}\over{\partial  r}}  =  &  -\displaystyle{1\over
 r^\alpha}{{\partial r^\alpha J_k}\over\partial r} + \dot R_kW_k\\ & P
 & = \displaystyle{{\rho RT}\over W}
\end{eqnarray}
where $\alpha=0$ for planar flames; $\alpha=1$ for cylindrical flames;
$\alpha=2$ for  spherical flames.  $\rho$ is the  density, $u$  is the
velocity, $P$ is  the pressure, $e$ is the  internal energy, $Y_k$ are
the species mass fractions for  $k=1,...,N$ species, $\dot R_k$ is the
molar rate of production of the $k^{th}$ species, $t$ is the time, $T$
is  the temperature,  $W$ is  the mean  molecular mass,  $W_k$  is the
molecular mass of the $k^{th}$ species, $J_\nu$ is the viscous flux of
momentum, $J_e$  is the  energy flux, $J_k$  is the  $k^{th}$ 
species diffusive flux, $J_n$ is the flux of heat and $R$ is the 
gas constant.\\

\noindent For all flame geometries ($\alpha=1,2,3$):\\

\noindent  $ \chi^\alpha=-\alpha  \left(  \displaystyle{{{2u}\over 3r}
{{\partial  \mu}\over {\partial  r}}} \right)$,  and  $\Gamma^\alpha =
\displaystyle{{1\over  r^\alpha}{\partial\over  {\partial r}}}  \left(
\displaystyle{   r^\alpha{{\lambda    RT}\over   C_v}   {\partial\over
{\partial r}} \left({1\over W}\right) } \right)$\\

\noindent  For planar  flames:\\ $  \Psi^0= \displaystyle{  {4\over 3}
({{\partial u}\over {\partial r}})^2 }$.\\

\noindent For cylindrical flames:\\ $\Psi^1 = \displaystyle{{4\over 3}
\left({{\partial u}\over{\partial  r}}- {u\over r}\right)^2  + {4\over
3} \left({u\over r} {{\partial u}\over{\partial r}} \right)} $\\

\noindent For spherical flames $\Psi^2$  has the same form as $\Psi^1$
except that the last term is absent.\\

Models for the various fluxes are given by,
\begin{eqnarray}
  J_\nu  &=&   -\displaystyle{4  \over  3}\mu  \displaystyle{{\partial
 u}\over  {\partial  r}}  \\  J_k  &=&  -\rho  D_k\left(\displaystyle{
 {{\partial   Y_k}\over  {\partial   r}}   -\displaystyle{Y_k\over  n}
 {{\partial  n}\over {\partial  r}} }\right)  +\rho  \Theta_kY_k +\rho
 V_cY_k  \qquad k=1,2,...,N   \\     
 J_e     &=&    -\displaystyle{\lambda\over     C_v}
 \displaystyle{{\partial   e}   \over    {\partial   r}}\\   J_n   &=&
 -\displaystyle{\lambda\over  C_v}  \displaystyle{{\partial  Y_k}\over
 {\partial r}}
\end{eqnarray}
$\mu$ is the dynamic viscosity, $\lambda$ is the thermal conductivity,
$n$ is the mole number, and $C_v$ is the specific heat capacity at 
constant volume.

Viscosities and binary diffusivities are evaluated using the theory of
Chapman  \&  Enskog (see \cite{REID1960}),  Thermal
conductivities  from \cite{MASON1962}.   Mixture
properties   are  evaluated from Wilkes formula (see \cite{REID1960}).   
The diffusive  fluxes  include the  Soret
thermal--diffusion for light species ($H$ and $H_2$).

Thus, all transport properties are evaluated locally and are functions
of the local temperature,  which  allows the system  to   be  strongly
inhomogeneous  in transport  properties and gives the simulations an
added degree of reality.  The  specific  Lewis numbers
are,  therefore, non-unity,  $Le_k =  Le_k(T)$, and  functions  of the
local temperature as well.

The same transport relations above are used in all of the
simulations, and flame speeds are calculated using an integral
definition, equation (4.1).

The  chemistry  is  modeled  as  a detailed  system  containing  $N$
chemical species, in $N_r$  elementary reactions, whose reaction rates
are given by,
\begin{eqnarray}
\dot  R_k  &=&  \sum_{j=1}^{N_r} (\nu^{''}_{kj}-\nu^{'}_{kj})  \left[{
k_j^f\prod_{l=1}^{N} \phi_l^{\nu^{'}_{jl}} - k_j^r\prod_{l=1}^{N}
\phi_l^{\nu^{''}_{jl}}}\right]
\end{eqnarray}
where  $\phi_l$ is the  molar concentration  of the  $l^{th}$ chemical
species,   and  $\nu^{'}$  and   $\nu^{''}$  are   the  stoichiometric
coefficients  of  reactants and  products  respectively.  The  forward
reaction rate constant  in reaction number $j$ is  given by Arrhenius'
law,
\begin{eqnarray}
k_j^f &=& A_jT^{\beta_j}\exp({-\Delta E_j/RT})
\end{eqnarray}
The  hydrogen/air  chemistry, shown  in  Table~1,  is a  comprehensive
system featuring 9 species and 30 reactions where the constants $A_j$,
$\beta_j$ and the  energy barrier $\Delta E_j$ were  obtained from
\cite{SUN2007},  with   some  modifications as noted in~\cite{MALIK2010}.
Thermodynamic data were computed using JANAF polynomials.



\protect \begin{table*} 
\begin{center}
{{\bf{Table  1.}} Detailed  H$_2$/O$_2$ Reaction  Mechanism  with rate
coefficients in  the form  $k=AT^{n}exp(-E/RT)$$^{\rm a}$ from  Sun et
al.~(2007). The rate proposed by Sutherland et al.~(1986) was used for
reaction (2).}

{\begin{tabular}{@{}cllcccccccccccc} \hline \\
\multicolumn{1}{c}{Number}& \multicolumn{2}{c}{Reaction} & $A^{\rm a}$
& $n^{\rm a}$  & $E^{\rm a}$ &  & & & & \\  \hline \\ [0.5ex] 1  & H +
O$_2$ $\rightleftharpoons$ O + OH & & 6.73E+12 &-0.50 & 69.75 & \\

2 & O + H$_2$ $\rightleftharpoons$ H + OH$^{\rm b}$ & & 5.08E+01 &2.67
& 26.33 & \\

3 &  H$_2$ + OH $\rightleftharpoons$ H$_2$O  + H & &  2.17E+05 &1.52 &
14.47 & \\

4 & Oh + OH $\rightleftharpoons$ O + H$_2$O & & 3.35E+01 &2.42 & -8.06
& \\

5 & H$_2$ + M $\rightleftharpoons$ H + H + M$^{\rm c}$ & & 2.23E+11 &0
& 40.196 & \\

  & H$_2$ + H$_2$$\rightleftharpoons$ H + H + H$_2$ & & 9.03E+11 & 0 &
  40.196 & \\
  
  & H$_2$  + N$_2$  $\rightleftharpoons$ H  + H +  N$_2$ &  & 4.58E+16
&-1.40 & 43.681 & \\
  
  & H$_2$  + H$_2$O $\rightleftharpoons$ H  + H + H$_2$O  & & 8.43E+16
&-1.10 & 43.681 & \\
  
6 & O  + O + M  $\rightleftharpoons$ O$_2$ + M$^{\rm d}$  & & 6.16E+09
&-0.5 & 0.0 & \\

7 & O + H + M $\rightleftharpoons$ OH + M$^{\rm d}$ & & 4.71E+12 &-1.0
& 0.0 & \\

8 & H + OH + M  $\rightleftharpoons$ H$_2$O + M$^{\rm e}$ & & 2.21E+16
&-2.0 & 0.0 & \\

9 &  O$_2$ +  H (+  M) $\rightleftharpoons$ HO$_2$  (+ M)$^{\rm  f}$ &
$k_0$ & 2.65E+13  &-1.30 & 0.0 &  \\ & & $k_\infty$ &  4.65E+09 &0.4 &
0.0 & \\

 & O$_2$  + H (+ H$_2$O) $\rightleftharpoons$  HO$_2$ (+ H$_2$O)$^{\rm
g}$& $k_0$ & 3.63E+13 &-1.00 & 0.0 & \\ & & $k_\infty$ & 4.65E+09 &0.4
& 0.0 & \\

10 & H$_2$ + O$_2$ $\rightleftharpoons$  HO$_2$ + H & & 7.40E+02 &2.43
& 223.85 & \\

11 & HO$_2$ + H $\rightleftharpoons$ OH + OH & & 6.00E+10 &0 & 1.23 &
\\

12 &  HO$_2$ +  O $\rightleftharpoons$ OH  + O$_2$  & & 1.63E+10  &0 &
-1.86 & \\

13  & HO$_2$  + OH  $\rightleftharpoons$ H$_2$O  + O$_2$$^{\rm  b}$& &
1.00E+10 & 0 & 0.00 & \\ & & & 5.80E+10 & 0 & 16.63 & \\

14 & HO$_2$ + HO$_2$ $\rightleftharpoons$ H$_2$O$_2$ + O$_2$$^{\rm b}$
& & 4.20E+11 &0 & 50.133 & \\ & & & 1.30E+08 & 0 & -6.817 & \\

15 & H$_2$O$_2$  (+ M) $\rightleftharpoons$ OH + OH  (+ M)$^{\rm h}$ &
$k_0$ &  1.2E+14 &  0 & 190.37  & \\  & & $k_\infty$  & 3.00E+14  &0 &
202.84 & \\

16 & H$_2$O$_2$ + H $\rightleftharpoons$ H$_2$O + OH & & 1.02E+10 &0 &
14.96 & \\

17 & H$_2$O$_2$  + H $\rightleftharpoons$ H$_2$ +  HO$_2$ & & 1.69E+09
&0 & 15.71 & \\

18 & H$_2$O$_2$ + O $\rightleftharpoons$ OH + HO$_2$ & & 8.43E+08 &0 &
16.61 & \\

19 & H$_2$O$_2$ + OH  $\rightleftharpoons$ H$_2$O + HO$_2$$^{\rm b}$ &
& 1.70E+15 &0 & 123.05 & \\ & & & 2.00E+09 & 0 & 1.79 & \\

20 & HO$_2$ + H $\rightleftharpoons$ H$_2$O + O & & 1.44E+09 &0 & 0 &
\\ \hline \\
\end{tabular}}
\end{center}

{$^{\rm a}$Units are kmole, m$^3$, s, K and KJ/mole.}

{$^{\rm b}$Reactions (13),  (14) and (19) are expressed  as the sum of
the two rate expressions. }

{$^{\rm c}$Chaperon efficiencies are  0.0 for H$_2$, H$_2$O, N$_2$ and
1.0 for all other species.}

{$^{\rm d}$Chaperon  efficiencies are 2.5  for H$_2$, 12.0  for H$_2$O
and 1.0 for all other species.}

{$^{\rm e}$Chaperon  efficiencies are 2.5  for H$_2$, 6.39  for H$_2$O
and 1.0 for all other species.}

{$^{\rm f}$Troe parameter  is F$_c$=0.57.  Chaperon efficiencies: 1.49
for H$_2$, 0.0 for H$_2$O and 1.0 for other species.}

{$^{\rm  g}$Troe parameter is  F$_c$=0.81. Chaperon  efficiencies: 1.0
for H$_2$O and 0.0 for other species.}

{$^{\rm  h}$Troe parameter is  F$_c$=0.50. Chaperon  efficiencies: 2.5
for H$_2$, 12.0 for H$_2$O and 1.0 for other species.}

\label{tab:reacsall}
\end{table*}


The pressure  field consists of  a mean and a  fluctuating components,
$p=\langle  p\rangle  +p'$;  similarly  for  the  velocity  $u=\langle
u\rangle +u'_p$.  Transmissive (non-reflecting) boundary conditions were
imposed on the  mean pressure and mean velocity  at the open (outer)
boundary, but the fluctuating quantities $p'$  and $u'_p$ were 
reflected back in to
the domain, which sustains the fluctuations inside the domain for long
periods. For  the curved flames we  have $u=0$ at  the centre ($r=0$).
For details of the simulation method and validation, see Malik and 
Lindstedt (2010, 2012)



In order to simulate conditions inside practical devices such as
engines, it is desirable to produce unforced pressure fluctuations
subject to natural decay due to the action of viscosity, which
is typical of conditions after ignition in an engine. This was
achieved by running a pre-calculation consisting of releasing
a set of square pressure ramps of large amplitude inside a domain
of length $L$; these pressure ramps then diffuse, convect and reflect
back from the ends of  the  domain  until random  pressure
fluctuations of small amplitude with a range of frequencies are
obtained. When the fluctuations have decayed down to  a suitable
level (about $2\%$ of atmospheric pressure), the main simulations
were started ($t=0$) by introducing the fuel
inhomogeneity  as a sinusoidal oscillation cycle  in the
equivalence ratio just  ahead of  the flame front,
$\phi(r)=1.0+A_\phi\sin(2\pi r/\lambda_\phi)$, where $\lambda_\phi$
is the length scale of variation and $A_\phi$ is the initial amplitude.
Four  cases   of  moderate inhomogeneity     $A_\phi\ll     1$     
were     consider,     $A_\phi =-0.2,-0.1,0.1,0.2$.

A domain size of $L=400mm$  was found to produce pressure oscillations
in  the required  frequency range  $200-1000$Hz which are known to 
couple to the flame thermo-chemistry.






\begin{figure}
  \centerline{\includegraphics[height=6.5cm]{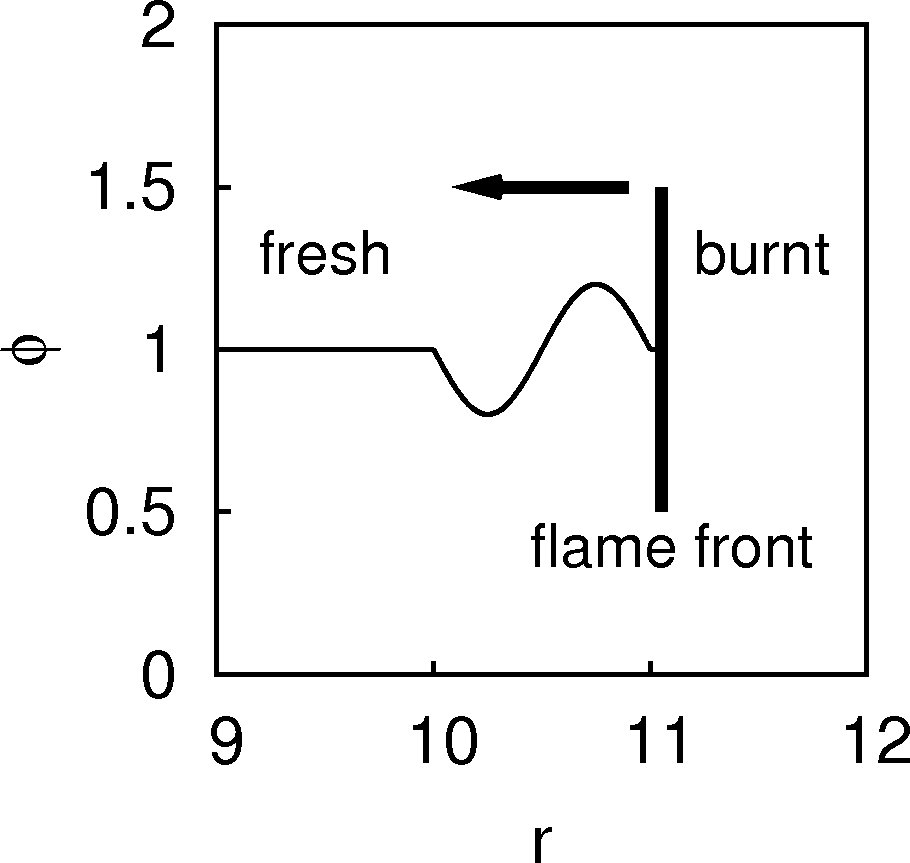}}
\caption{Schematic  of  the  start  of the  simulations,  showing  the
equivalence ratio $\phi$ against the  radius $r$. The flame propagates 
from right to   left  into  the   inhomogeneous  fuel   distribution.   
(Pressure fluctuations are also active in the domain)}
\label{fig:F2}
\end{figure}

Figure~1 shows a schematic of the situation just as the flame is about
to propagate  into the  equivalence ratio oscillation  cycle. Figure~2
shows  typical  snapshots  of  (a)  the pressure  field  and  (b)  the
associated velocity field across  the H2/air cylindrical flame at some
instant.  The non-zero velocity observed ahead of the flame is induced
by the  pressure fluctuations which exist throughout  the domain.  The
location of the flame front  is indicated by the vertical dashed lines
in Figure~2  -- one can  observe the jump  in velocity due to  the gas
expansion behind the flame.\\



\begin{figure}
\centering
\hspace{0cm}
\mbox{\hspace{-1.3cm}
      \subfigure{\includegraphics[height=6cm]{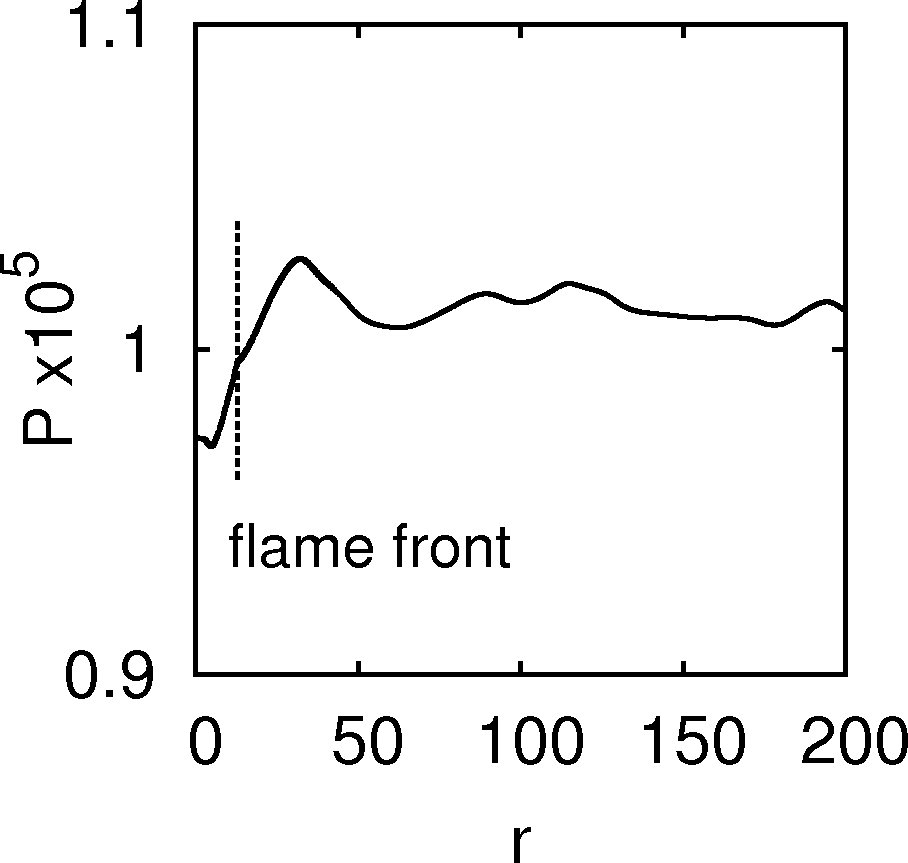}} \hspace{0cm}
      \subfigure{\includegraphics[height=6cm]{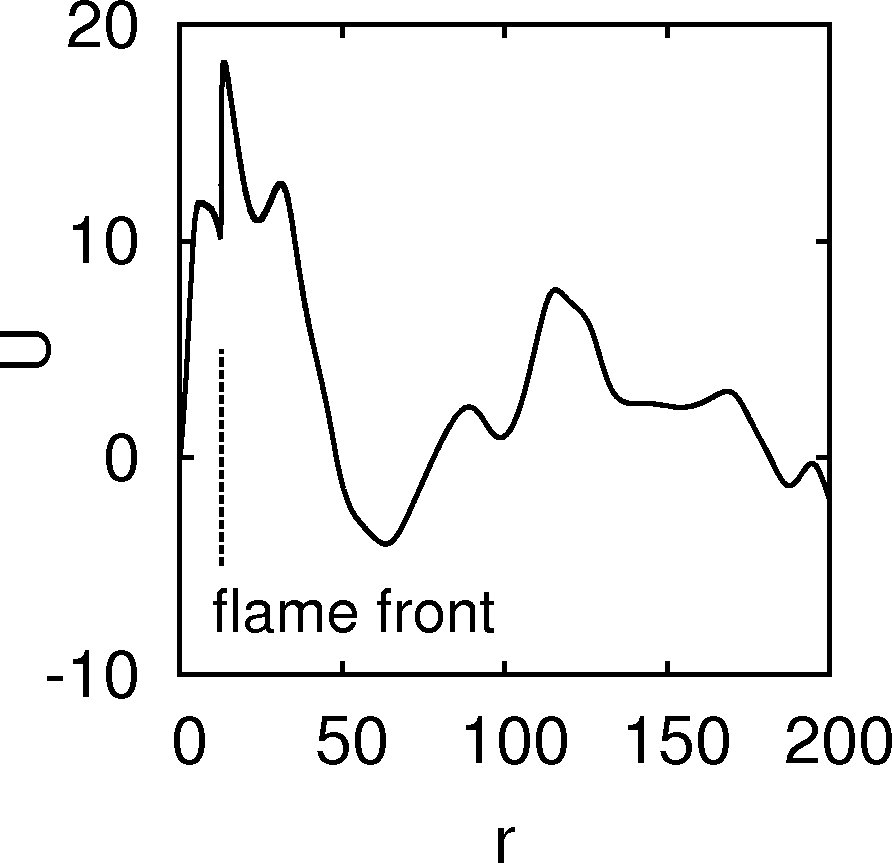}} }
\caption{{\em Left:} The pressure  field and, {\em Right:} the 
velocity field  across a $200$~mm section of the domain  taken  from a  
typical  simulation  of a cylindrical flame. The vertical  dashed lines 
indicate the location of the flame front} 
\label{fig:F3}
\end{figure}



\begin{figure}
\centering
\mbox{\subfigure{\includegraphics[height=6.5cm]{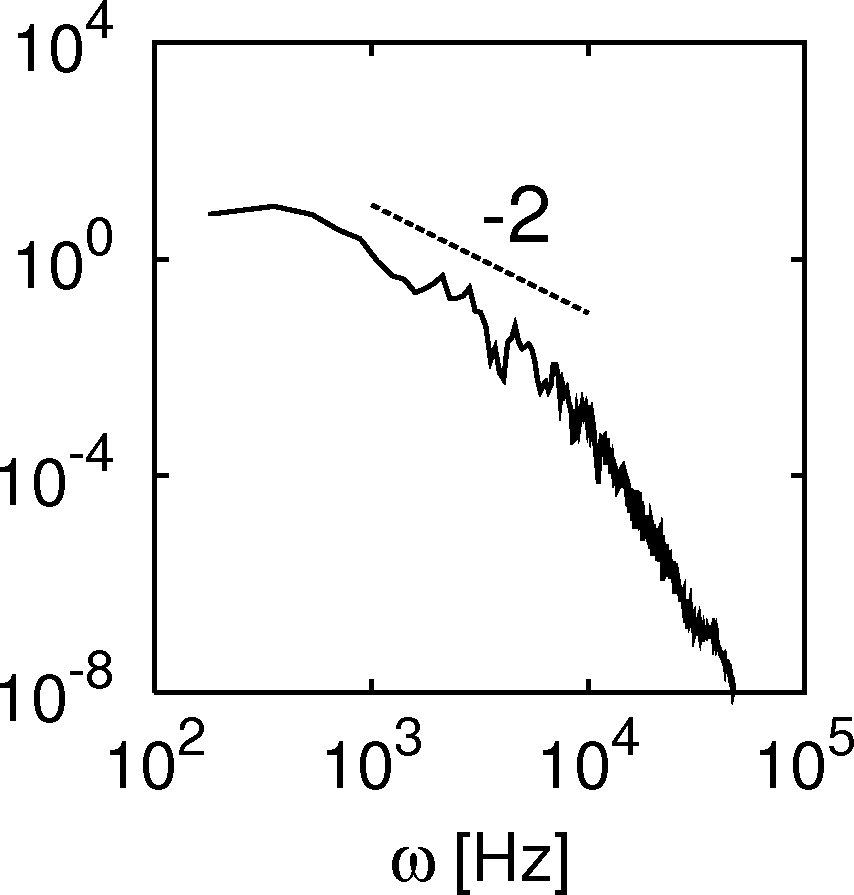}} \hspace{0cm}
           \subfigure{\includegraphics[height=6.5cm]{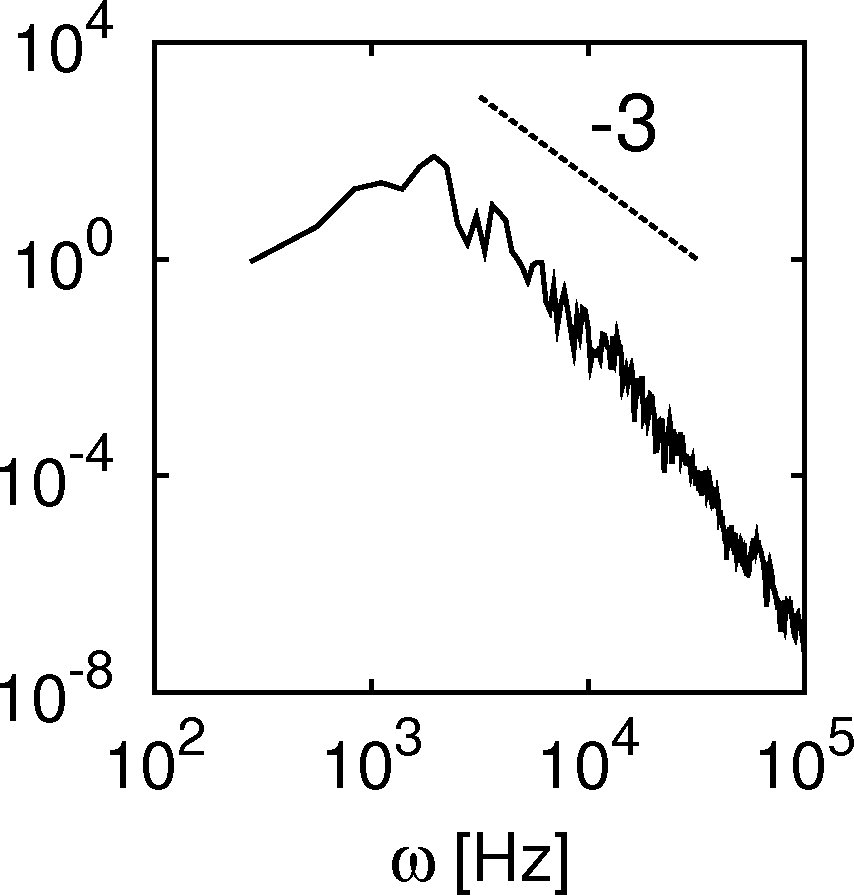}} }\\
\mbox{\subfigure{\includegraphics[height=6.5cm]{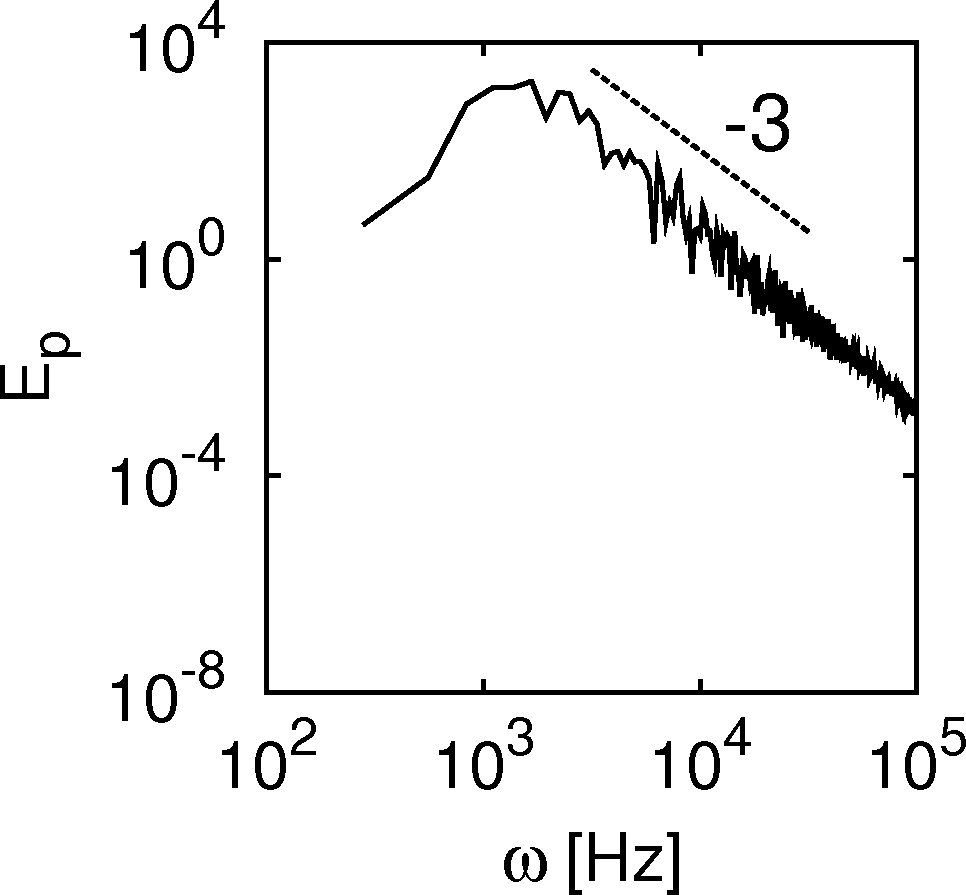}} }
\caption{Log--log spectra  of pressure fluctuations  $E_p$, normalised
by $E_p(0)$,  against the frequency  $\omega$ [Hz]: {\em top left:}  
Planar flame, {\em top right:}  Cylindrical flame, {\em bottom:}  
Spherical flame}
\label{fig:F4}
\end{figure}

\section{Results}

\subsection{Pressure Spectra}

Pressure signals were
recorded at points close to the flame fronts, spaced sufficiently far so as to avoid
duplicate signals, yielding a total of five peridograms from which the final spectra
were obtained by averaging. The signals were initially subjected to smoothing using a cubic spline solver on to 4096 point signal.
Figure~3  shows the  pressure spectra  $E_p(\omega)$ in  log-log scale
from the flame simulations: 3(b) from the cylindrical flame,  and 3(c) 
from the spherical flame, and for comparison the spectrum from the  
planar flame figure~3(a) from~\cite{MALIK2010} is
also included. All spectra are normalised by $E_p(0)$.  
The spectra is an indicator of  the strength of flame response to pressure
fluctuations.   Stronger flame-pressure  interaction should  produce a
shallower gradient  in the spectral  broadening.  In the case  of high
frequency pressure  forcing which are  known not to interact  with the
flame  kernel,  the spectrum  falls  off  very  steeply (as  seen  in
Figure~4 in \cite{MALIK2010}).\\



\begin{figure}
\centering
\mbox{\hspace{-1cm}
      \subfigure{\includegraphics[height=6.5cm]{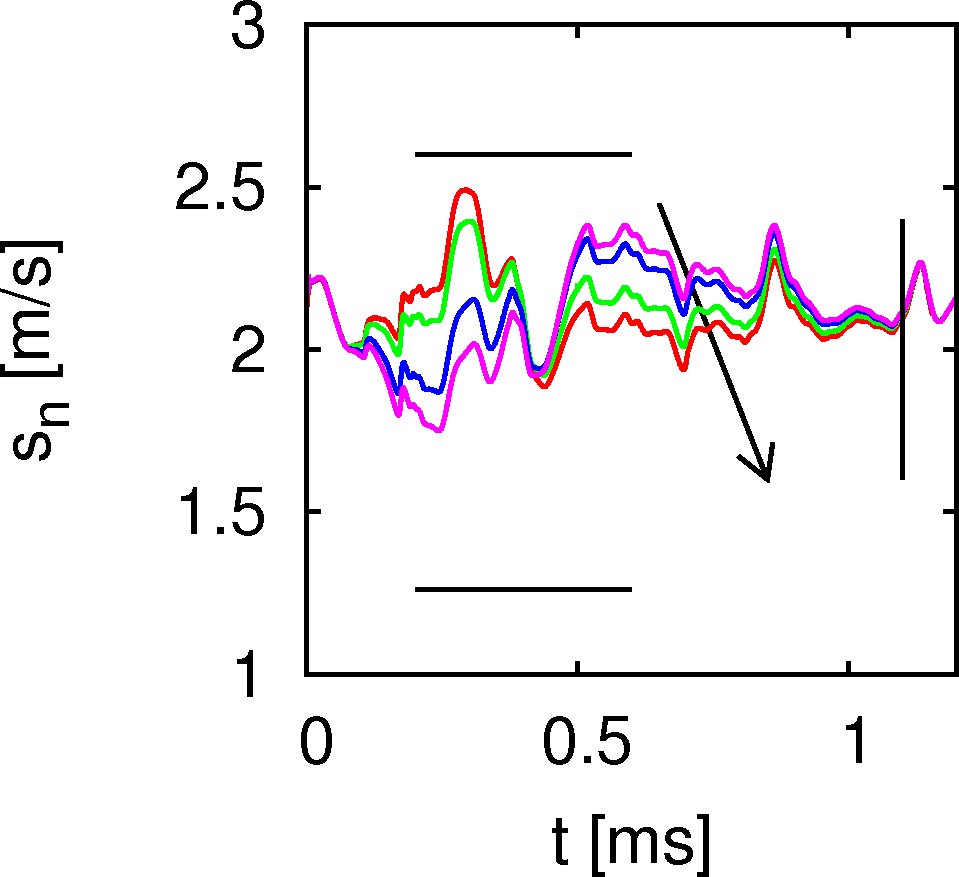}} \hspace{0cm}
      \subfigure{\includegraphics[height=6.5cm]{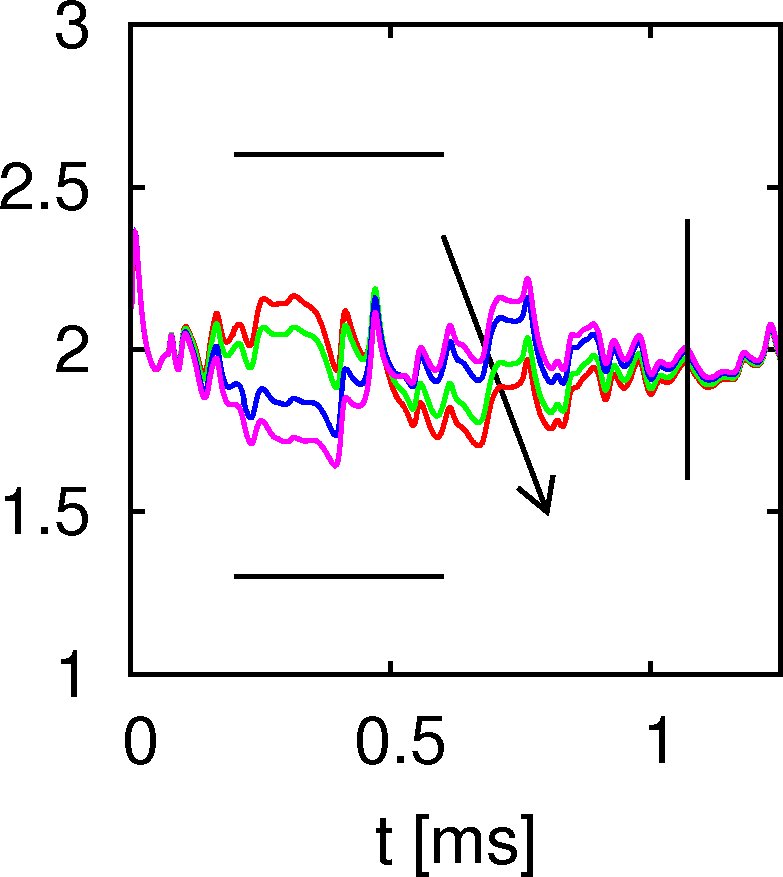}} }\\
\mbox{\subfigure{\includegraphics[height=6.5cm]{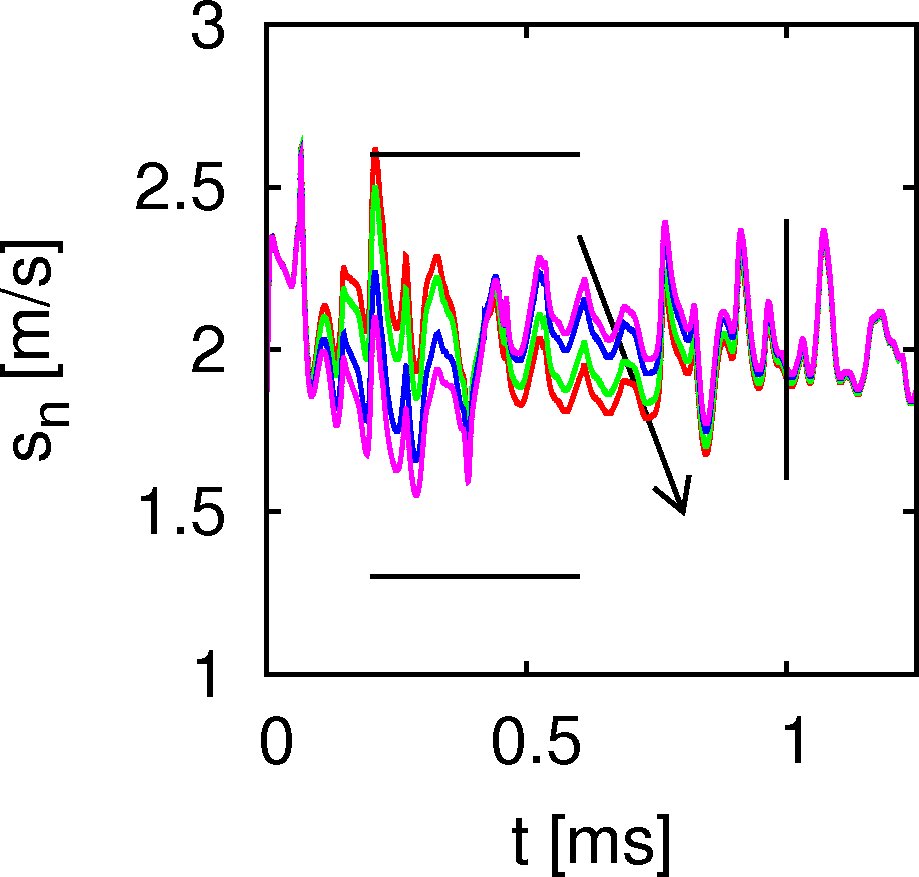}} }
\caption{The burning  velocity $s_n$ [m/s] against time  $t$ [ms]: 
{\em top left,} Planar flame;  {\em top right,} Cylindrical flame; 
{\em bottom,} Spherical  flame.  Four cases of equivalence  ratio 
amplitude are  shown in each figure, the thick arrow
showing the direction of increasing amplitude: $A_\phi=0.2$
(red), $0.1$ (green), $-0.1$  (blue), $-0.2$ (purple).  The theoretical
bounds for stationary conditions at $\phi=1.2$ and $0.8$ are indicated
by  the  horizontal  lines. The  vertical  lines indicate  the times 
when the curves come together as they leave the fuel oscillation cycle.}
\label{fig:F5}
\end{figure}

In the pressure spectra from the inwardly propagating flames, Figures~3(b) 
and 3(c), the lower frequencies $\omega<1000Hz$ are  the forcing pressure 
frequencies (similar to Figure 3(a) from the planar flame). But at higher frequencies, $\omega>1000Hz$,  the spectra in Figures~3(b)  and 3(c)  
scale  close to  $\sim  \omega^{-3}$ which  is steeper than  in the 
expanding expanding  H2/air flames spectra,  figure 3 in \cite{MALIK2010}, 
which scale 
close to  $\sim \omega^{-2}$  over shorter  ranges centred around  
$\omega\approx 5000 Hz$.  In  the current  simulations,  the 
cylindrical flame spectrum in Figure~4(b) shows  a steeper fall off 
from $\omega
\approx 10,000 Hz$, but the spherical flame spectra, Figure~4(c),
continues to scale like $\sim \omega^{-3}$ even at very high frequencies.

The  steeper  spectral   scaling  indicates  a  weaker  flame-pressure
interaction  than in  the expanding flames although the excitations appear 
over a wider  range of frequencies. Interestingly, the corresponding 
spectra from the expanding methane/air flames, figure 3  in \cite{MALIK2012a},
also scale close to $\sim \omega^{-3}$. 

As methane/air is a weakly
reactive fuel compared to hydrogen/air, this indicates that negative 
curvature may be responsible for dampening flame-pressure interactions. 
Alternatively, as there is no strain ahead of the flame in contracting flames,
then a reduction of strain-pressure interaction could also be responsible
for the steeper fall off in the spectra. This is consistent with
the expansing methane/air flame where the gas velocity is much smaller
than in  H2/air flames, resulting in low strain rates and overall weaker
strain-pressure interactions.

\subsection{Burning Velocity $s_n$}



\begin{figure}
\centering
\mbox{\hspace{-1cm}
      \subfigure{\includegraphics[height=4.5cm]{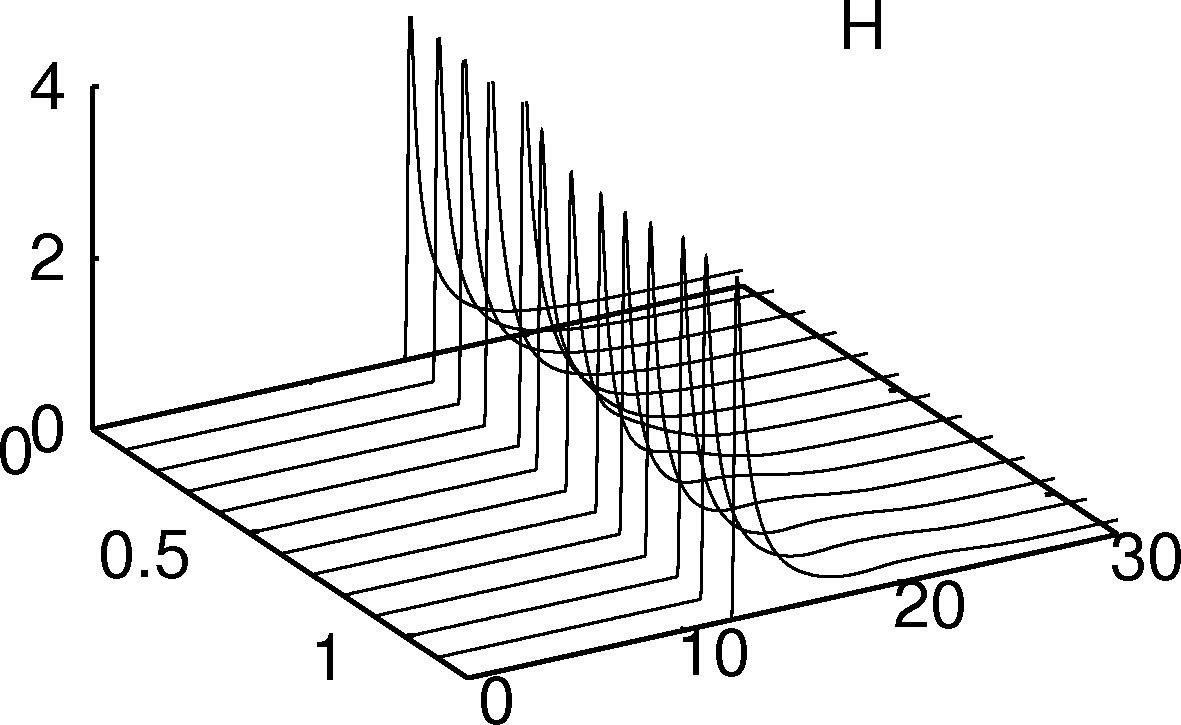}} \hspace{-0cm}
      \subfigure{\includegraphics[height=4.5cm]{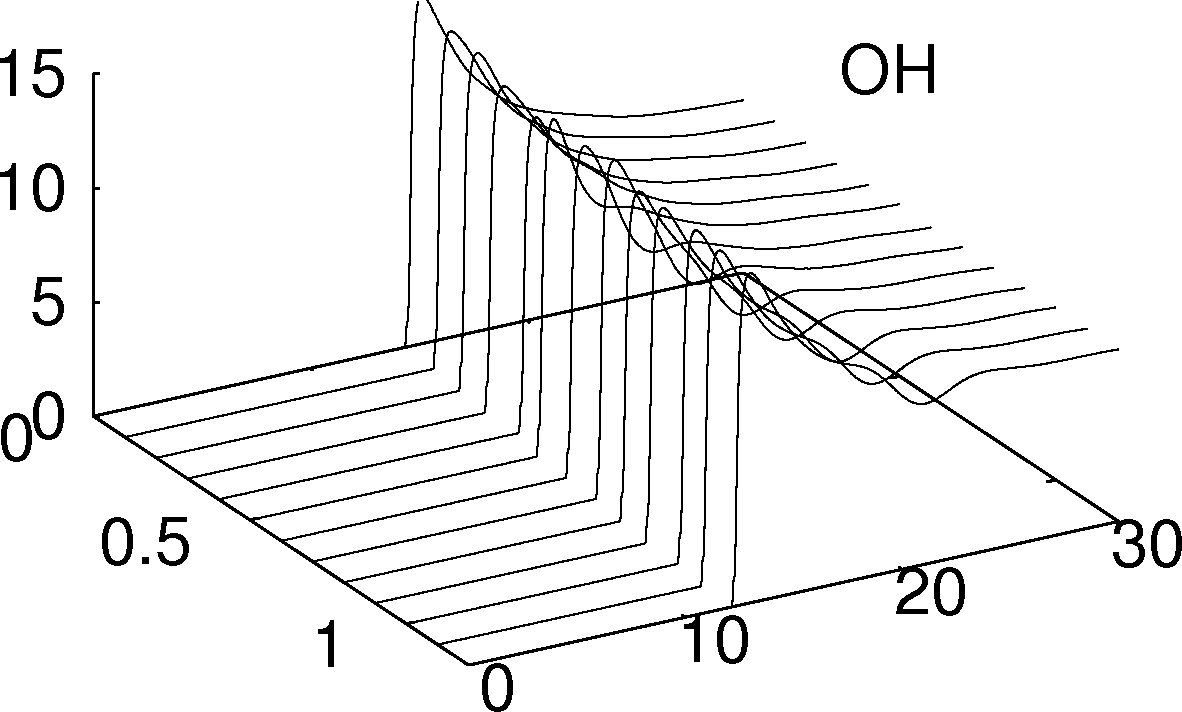}} } \\
\mbox{\hspace{-1cm}
      \subfigure{\includegraphics[height=4.5cm]{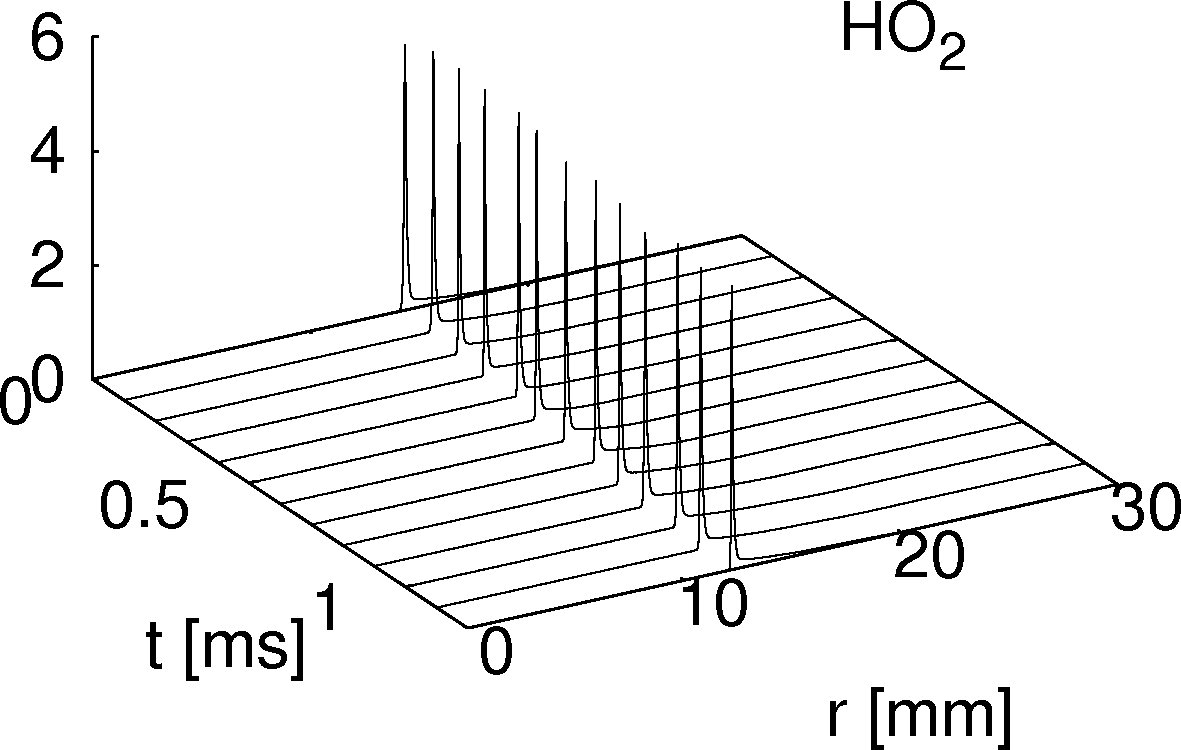}} \hspace{-0cm}
      \subfigure{\includegraphics[height=4.5cm]{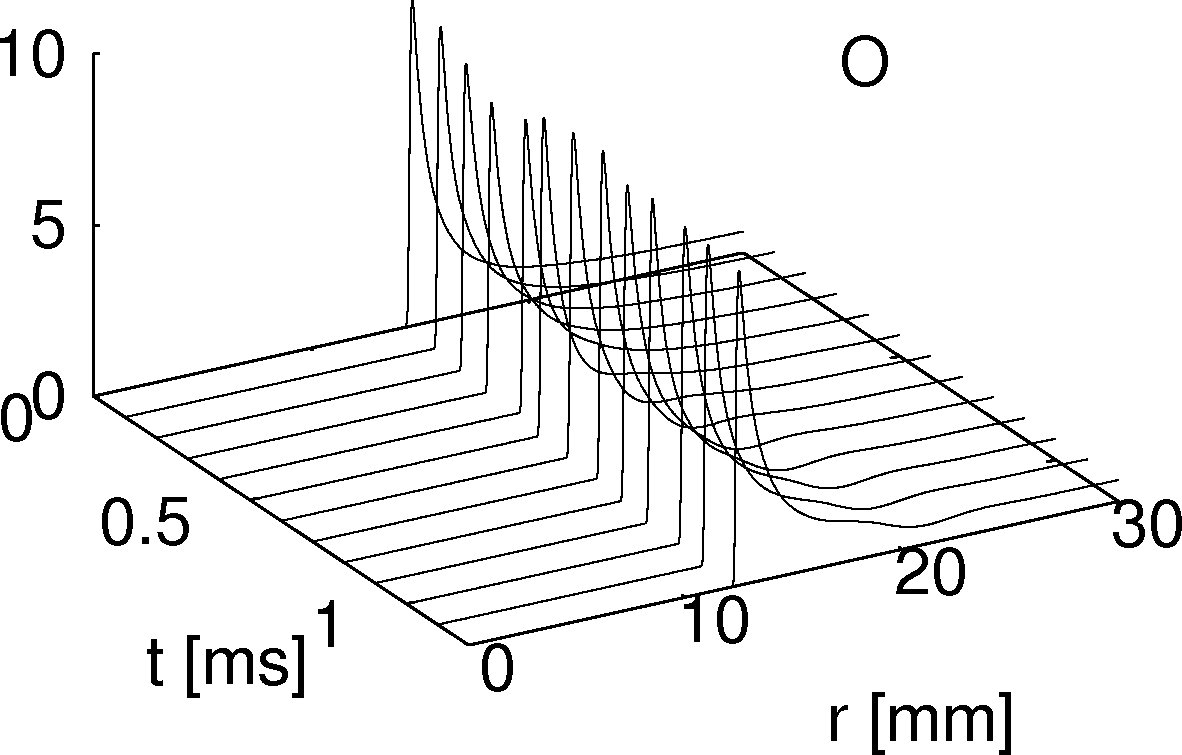}} }
\caption{Mole--fractions  of  key  chemical  species, as indicated, at
$0.05$~ms intervals, against time $t$~[ms] and radius $r$~[mm], from a
cylindrical flame with $A_\phi=0.2$}
\label{fig:F6}
\end{figure}

We  define  the  flame  speed   (burning  velocity)  as  the  rate  of
consumption of the fuel integrated across the flame,
\begin{eqnarray} 
  s_n &=& {1\over\rho_u (Y_F^u-Y_F^b)}
\int_{-\infty}^{+\infty}\dot{R}_FM_F dr
\end{eqnarray}
where $F$  is the fuel ($H_2$),  $M_F$ is the molecular  weight of the
fuel and $\rho_u$ is the density of the unburnt gases.  $Y_F^u$ is the
unburned  fuel mass  fraction  and  $Y_F^b$ is  the  burned fuel  mass
fraction  (which  is  close   to  zero  for  stoichiometric  and  lean
mixtures).

Figure~4 shows  the flame speed $s_n [m/s]$  against time $t [s]$   as 
the flames
propagate through the equivalence  ratio oscillation cycle.  The times
when the flames  leave the oscillation cycle are indicated by the 
vertical dashed lines.

Two trends are evident in  Figure~4.  First, the variation in the mean
due to the underlying sinusoidal fuel  distribution can be seen 
especially if one follows in time any one of the  cases in the figures, 
say the red line corresponding  to  $A_\phi=0.2$.   As  this  is  an  
unsteady  decaying system,
molecular transport processes compete  with the flame propagation from
the  start and  partially smoothens out the  fuel distribution  as the
flame moves through the fuel inhomogeneity.  The effect is an 
approximately sinusoidal but decaying variation in the burning velocity 
$s_n$.

Second,  superimposed on  the mean trend  are  the   higher  frequency
fluctuations in $s_n$ induced by the local flame-pressure interactions.  
As in the   outwardly  propagating  hydrogen/air   flames in  
\cite{MALIK2010}
periodic pressure fluctuations reflecting  back from the open boundary
and  impacting on  the reaction  layer  is evident  especially in  the
spherical  flame where  the  focusing of  pressure  waves towards  the
centre is the strongest.

The flame  speed $s_n$ does  not approach the limits  corresponding to
the  upper and  lower  bounds  of the  initial  fuel distributions  at
$\phi=1.2$   and  $\phi=0.8$  respectively,   indicated  by   the  two
horizontal  lines  in  Figure~4.   This is  partially due  to the  
smoothening  out   of  inhomogeneities  due   to  action of molecular  
transport processes; this is an unsteady system and by the time  
that the flame arrives at the point where $\phi$ was initially at its 
peak value of $1.2$, the local value of $\phi$ is
now closer  to the mean value of  $1$, i.e. $1< \phi  < 1.2$.  Similar
considerations apply to the lower bound where now 
$0.8<\phi<1$. (However,  we would expect the
upper and lower bounds to  be approached if the fuel distribution were
stationary which would be possible if the molecular diffusion were 
very small.)   

It is also possible that the flame response to the local rate of heat 
release is affected by a memory effect which could also prevent the 
flame speed from attaining to its extreme values. Memory effects are
discussed further in section 4.5.

\subsection{Relaxation number $n_R$}

The relaxation time  $\tau_R$ is the time it takes  the flame speed to
return to the mean value once disturbed. From Figure~4 $\tau_R$ is the 
time after which the different  $s_n$-curves come to within 5\% of the 
mean value $\phi=1$ and remain together thereafter. These times are 
indicated by the vertical lines in Figure 4.

A non-dimensional relaxation number $n_R$ is defined, \cite{MALIK2012a},
as
\begin{eqnarray}
  n_R &=& {\tau_R \over \tau_L}
\end{eqnarray}
which  is the  ratio of  the  relaxation time  to a  flame time  scale
$\tau_L$.  $n_R$ is a function of stretch, molecular transport
processes, the  fuel distribution, and  possibly also of  the pressure
fluctuations.  In general, we expect $n_R$ to  be a function of many 
parameters including Schmidt number,  Prandtl number and Lewis  numbers.   
For example  for small Schmidt   numbers, the smoothening  of the  fuel   
distribution
inhomogeneities  due  to the  stronger  molecular  diffusion could  be
more rapid leading to smaller relaxation numbers.

In  \cite{MALIK2010} it  was noted  how effectively  positive stretch
reduces  the relaxation  times $\tau_R$  by about  40\%:  from 
$\sim 1.1$~ms (planar), to $\sim 0.8$~ms (cylindrical); and $\sim 0.65$~ms (spherical). In  contrast,  from  Figure~4  negative curvature  
reduces $\tau_R$ by only about 10\% from the stretch-free planar flame
case:  we  have  $\tau_R  \sim  1.07ms$ in  the  inwardly  propagating
cylindrical flame,  and $\sim1.0ms$ in  the spherical flame case.

However, the non-dimensional $n_R$ from both the expanding H2/air and 
expanding CH4/air flames were found to  be similar and to follow the 
same trend:  $n_R$  reduces by about 30\% going from planar to 
spherically CH4/air flames and about 40\% in the H2/air flames.
Table~2 shows the relaxation times $\tau_R$ and the relaxation 
numbers $n_R$ from the present imploding H2/air flames, and also from
the two expanding flames (from \cite{MALIK2012a}) for comparison.

In defining $n_R$ from equation (4.2), we have used $\tau_L\sim
\delta_L/s_L$ and the flame  thickness $\delta_L$ is taken  as the  
width of the fuel reaction rate profile $\dot R_F$, as discussed in
\cite{MALIK2012a}. From the current implosion simulations we  obtain  
$n_R\approx 4.5$  in the  cylindrical flame,  and $n_R\approx 4.2$  
in the  spherical flame.


\protect \begin{table*} 
\begin{center}

\begin{tabular}{*{8}{l}}

\cr \hline \cr
   &$\Large \tau_R$  [ms]  &&  &$\Large n_R$  &&  \cr
   Propagation Mode  & Planar  &
   Cylindrical &  Spherical & Planar & Cylindrical &  Spherical &
   Reference \cr &&&&&&  \cr \hline \cr
   H$_2$/air  imploding &$1.1$ &$1.07$ &$1.00$ &$4.6$ &$4.5$ &$4.2$ &
   Current \cr

&&&&&&& \cr 
   H$_2$/air  expanding &$1.1$ &$0.80$ &$0.65$ &$4.6$ &$3.3$ &$2.7$ & a\cr

&&&&&&& \cr 
   CH$_4$/air expanding &$6.0$ &$4.50$ &$4.00$ &$4.4$ &$3.3$ &$3.0$ & b\cr
\hline

\end{tabular}


{\bf Table 2.} Relaxation times  $\tau_R$  and the  relaxation
numbers $n_R$ for hydrogen--air and methane--air flames.
{\em References:} a -- \cite{MALIK2010}; b -- \cite{MALIK2012a}

\end{center}
\end{table*}

The results shown in Table~2  are consistent with the spectral results
observed in  section~4.1 where negative stretch  in imploding flames
appears to  interact more weakly  with the pressure  fluctuations than
with positive  stretch in  expanding flames. The  weaker interaction
manifests itself here  as a smaller reduction in  the relaxation number
$n_R$.

\subsection{Unsteady thermochemical flame structure}

The  response of the  thermochemical flame  structure to  the unsteady
processes  is shown in  Figure~5 where  the mole-fraction  profiles of
selected species against time and  radius at 0.1ms intervals are shown
from the cylindrical flame (spherical flames produce qualitatively 
similar results). 

Figures~6 and 7 show the  mole-fraction profiles against the radius of 
all  the  species (except   inert  nitrogen) in,  respectively, the 
cylindrical and spherical flames at 0.125ms intervals.

The stable species, such as  $H_2$ and $H_2O$, show little response to
the oscillations most probably because they are already `saturated' at
high   concentration  levels   and  so   small  oscillations   in  the
mole-fractions produces little effect.

The observed response of the thermochemical structure in these figures
is similar to that found in the positively stretched flames in
\cite{MALIK2010, MALIK2012a}.  A {\em spectrum} of chemical species 
thickness $\{l_k\}$ rather  than a  single global  flame thickness  
$\delta_L$  seems more appropriate in characterising flame structure.

We may  reasonably extend  this idea to  fluctuations in  any physical
variable  that alters  the flame structure, and suggest the  existence 
of  an associated spectrum of non-dimensionalised length scales 
$\{S_k\}$,
\begin{eqnarray} 
   S_k &=& {l_k\over \lambda}\qquad k=1,2,...,N
\end{eqnarray}  
where $\lambda$  is  a characteristic length  scale  of  the  physical 
fluctuations. Thin profiles of those species such that $S_k\ll 1$ will 
not be significantly affected  except in amplitude --- such as $HO_2$
and $H_2O_2$; but the thicker profiles of those species with $S_k\gg
1$  will also  be perturbed  across  their whole  length because  such
oscillations  can  penetrate  inside  their  structure -- such as $OH$;  
while  stable species, such as $H_2$, $O_2$ and $H_2O$  are likely  to  
be  unaffected for  the  reason noted  above. Other species display 
intermediate stages of profile disturbance depending on the thickness
of their profile.\\



\begin{figure}
  \centerline{\qquad\includegraphics[width=15cm,clip]{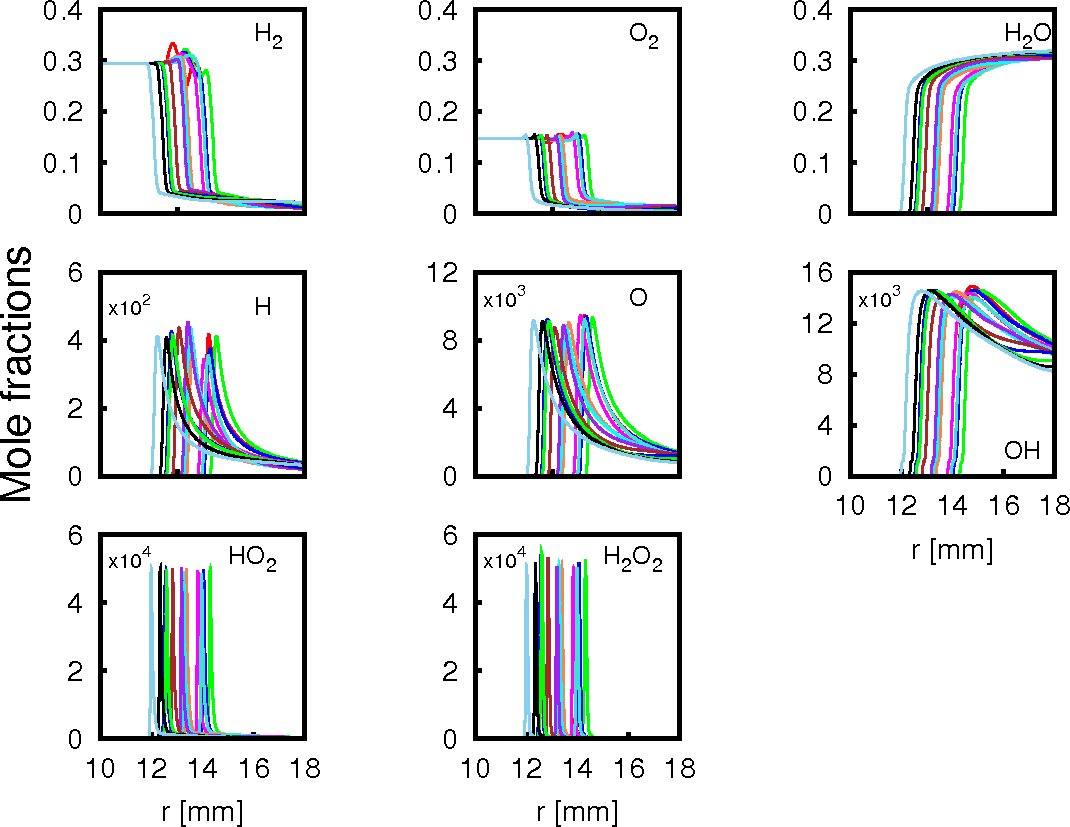}}
  \caption{Mole--fractions  $X_k, \quad k=1,2,...,8$ of all chemical  
species (except Nitrogen), as  indicated on each  plot,  at  $0.1$~ms  
intervals  against the  radius $r [mm]$,  from  the cylindrical 
flame with $A_\phi=0.2$. The mean flame propagation is from right to 
left (the centre).}
\label{fig:F7}
\end{figure}

%


 
\begin{figure}
  \centerline{\qquad\includegraphics[width=15cm,clip]{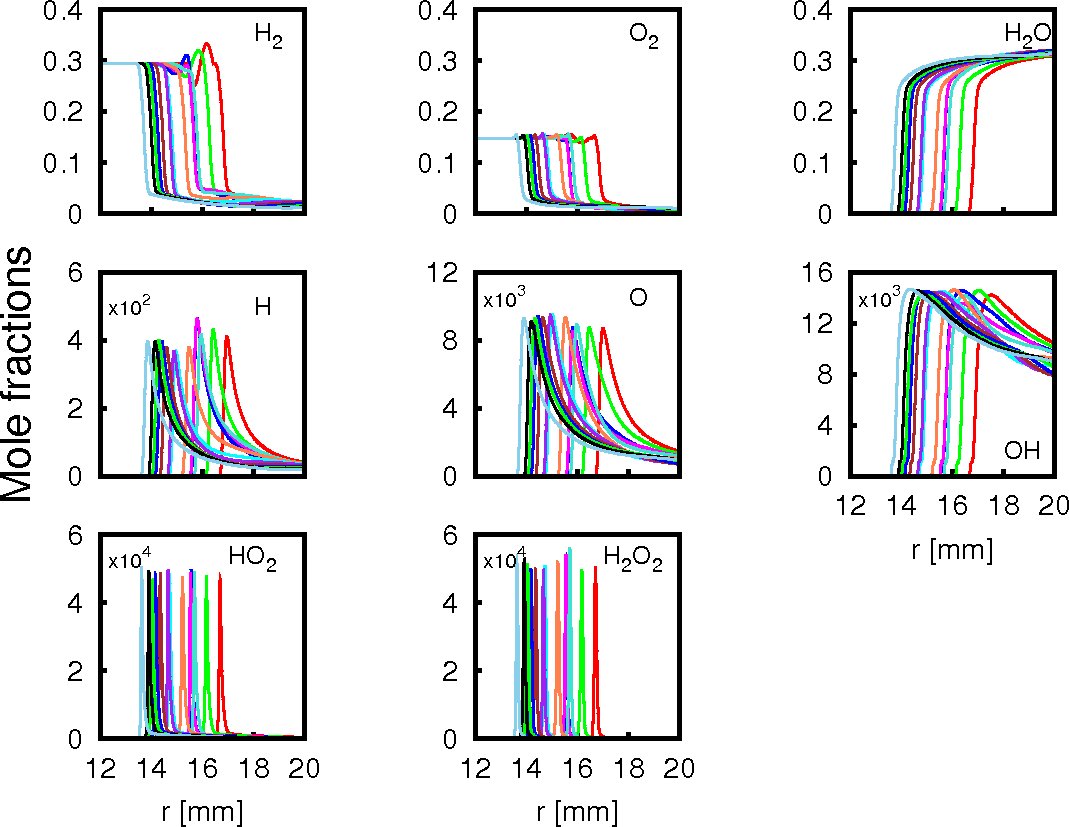}}
  \caption{As Figure 6, but for the spherical flame}
\label{fig:F8}
\end{figure}

\subsection{Time-lag and `flapping'}

An issue  of practical and theoretical  interest is the  impact of the
physical oscillations on the  chemical balance and heat release inside
the  disturbed flame  kernel  and the  consequent  induced time--lag.

Figure~8 shows the profiles at 0.1ms intervals, from the cylindrical 
flame simulations with $A_\phi=0.2$, of (a)  the rate of heat release 
$\dot h$, (b) the mole-fractions of  H-radical, and  (c) the 
mole-fractions of  O-radical.  Figure~9 shows similar profiles from the spherical flames.


\begin{figure}
\centering
\mbox{ \hspace{-2cm}
            \subfigure{\includegraphics[height=5.5cm]{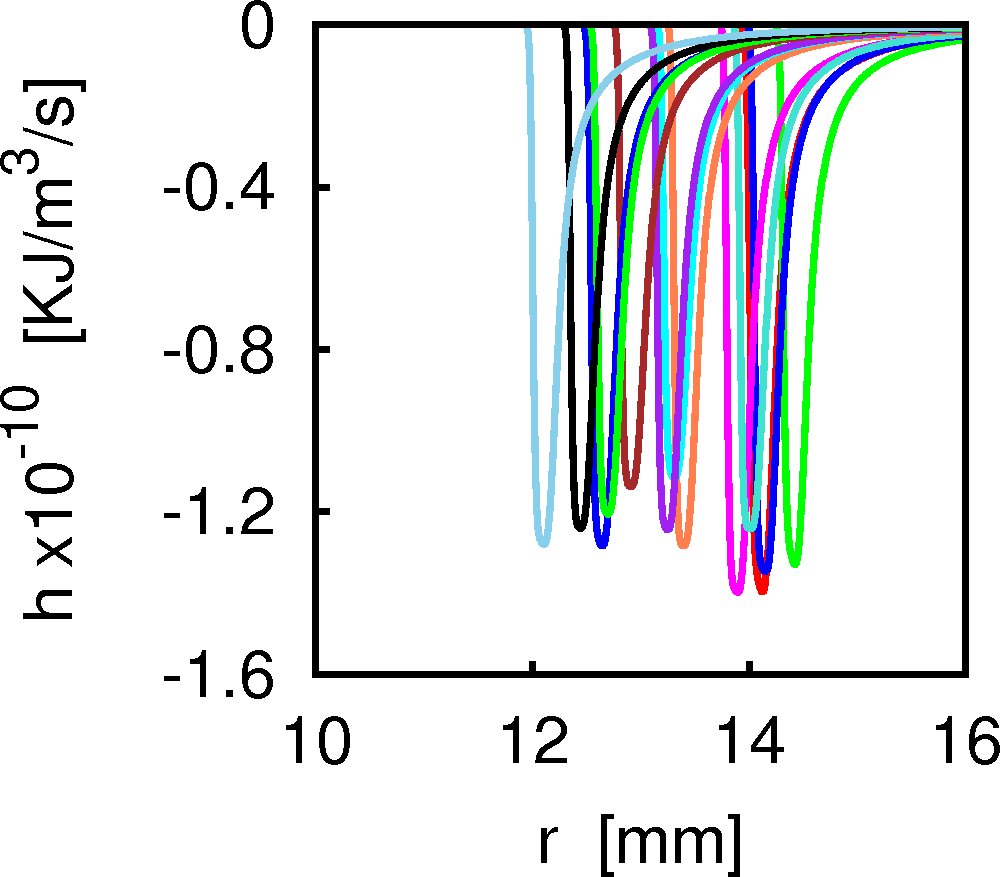}} } \\
\mbox{ \hspace{-0.5cm}
      \subfigure{\includegraphics[height=5.5cm]{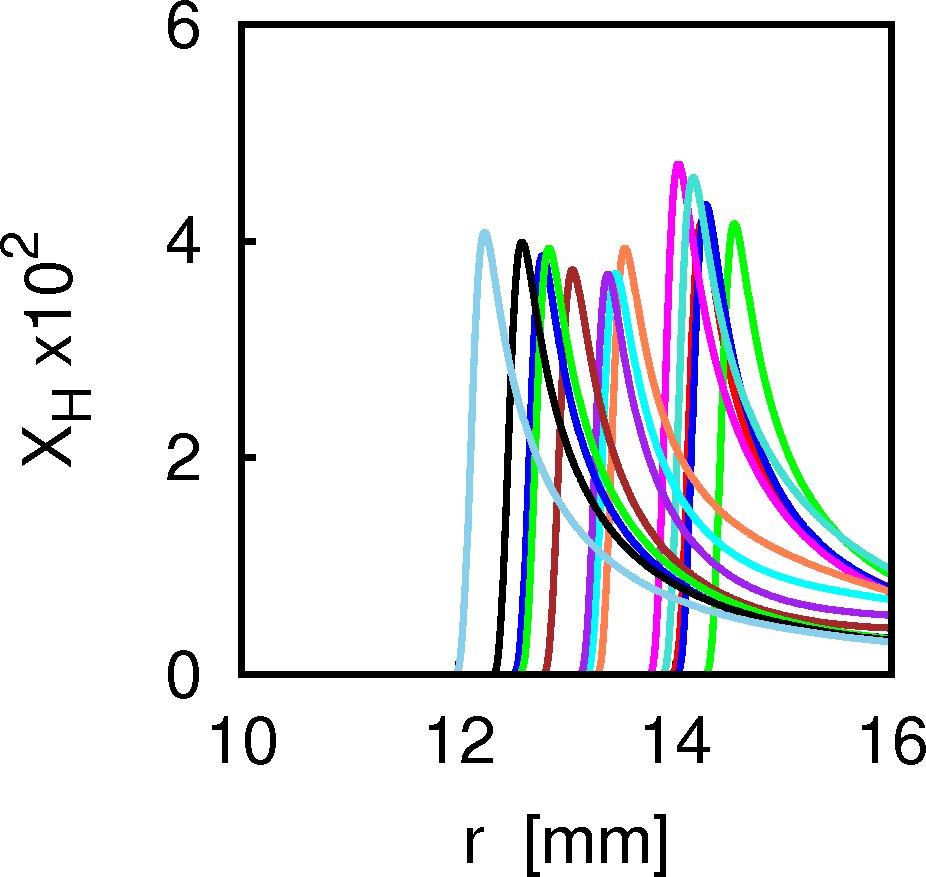}} \hspace{0.5cm}
      \subfigure{\includegraphics[height=5.5cm]{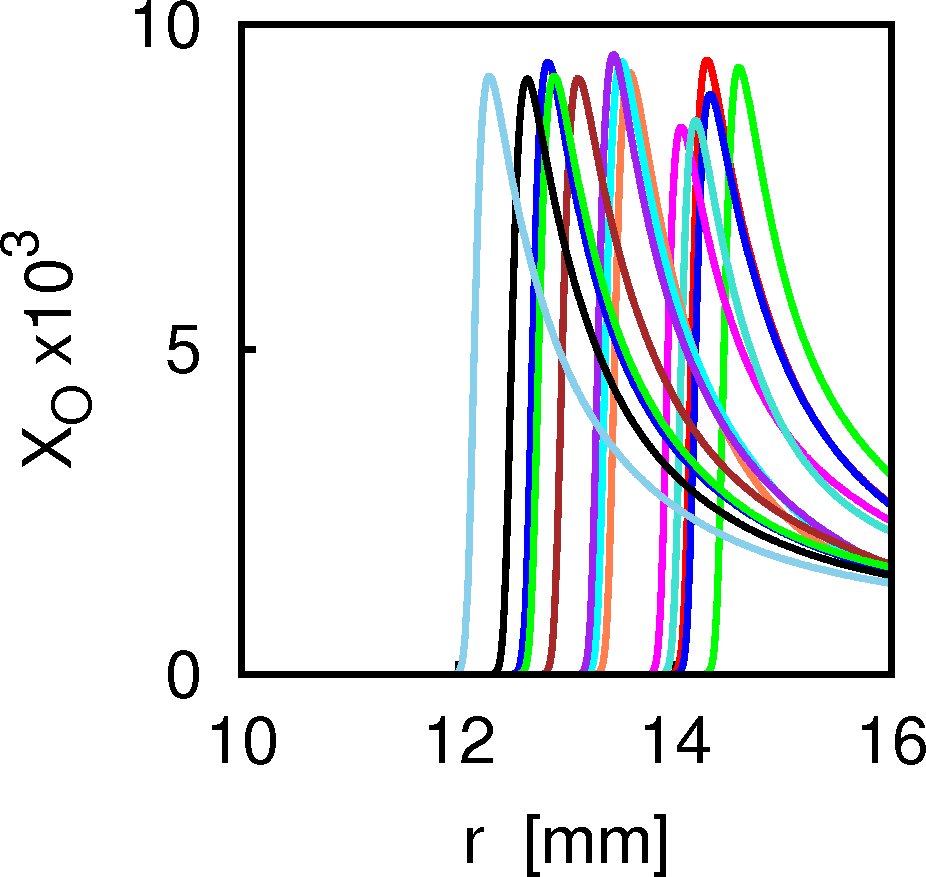}} }
\caption{{\em Top:} Rate   of  heat  release   per  unit  volume   
$\dot  h$ $[KJ/m^3/s]$ against the radius $r$  [mm] at 0.1ms intervals, 
from the cylindrical flame with  $A_\phi=0.2$. Corresponding 
mole-fractions of, {\em Bottom left:} H-radicals, 
{\em Bottom left:} O-radicals. The mean flame propagation is from right 
to left (the centre).}
\label{fig:F8}
\end{figure}

%


\begin{figure}
\centering
\mbox{ \hspace{-2cm}
            \subfigure{\includegraphics[height=5.5cm]{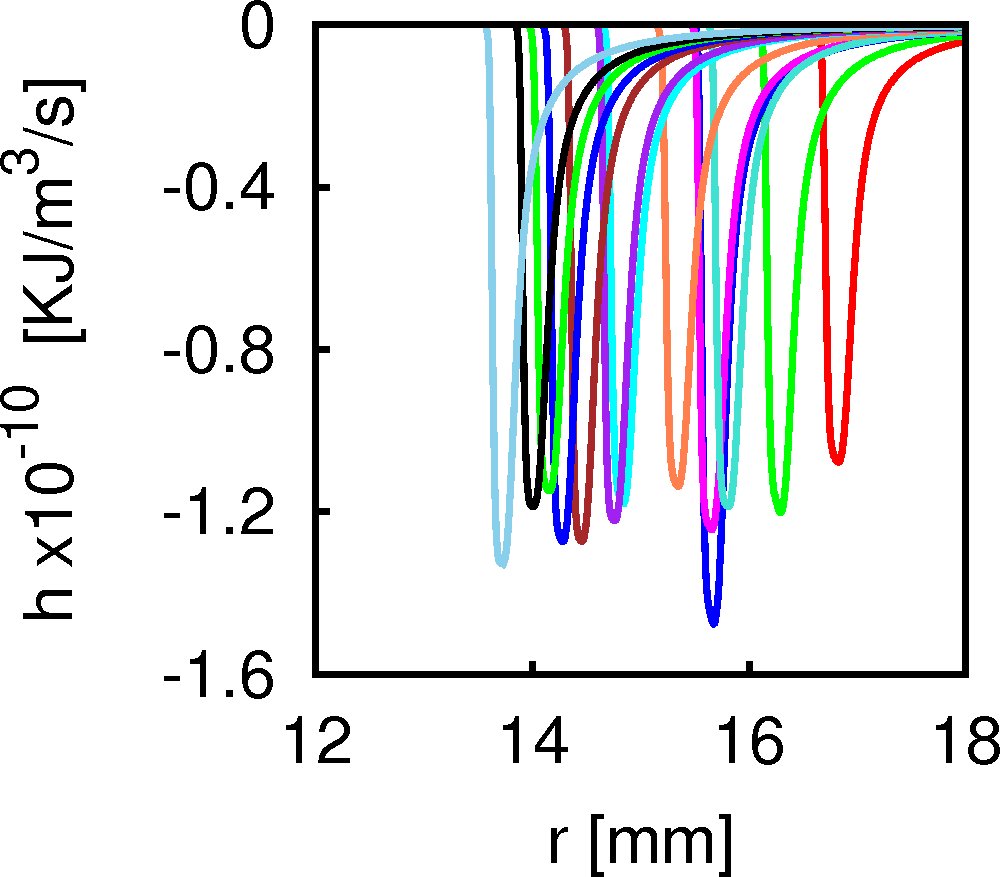}} } \\
\mbox{ \hspace{-0.5cm}
      \subfigure{\includegraphics[height=5.5cm]{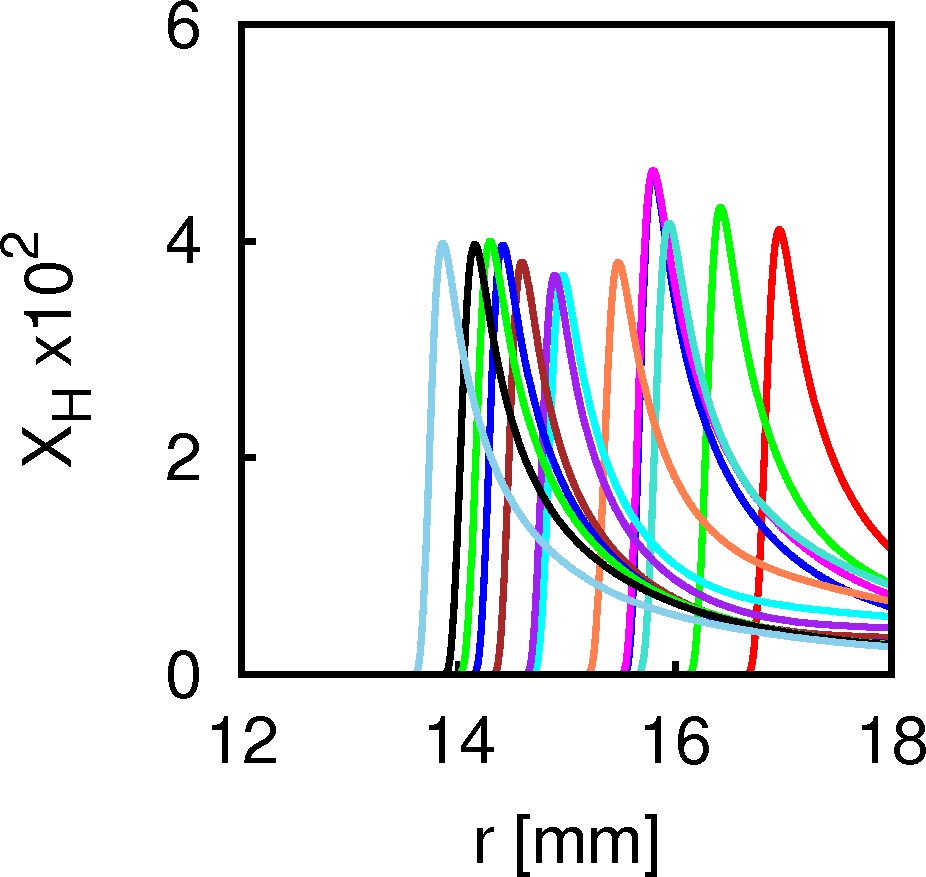}} \hspace{0.5cm}
      \subfigure{\includegraphics[height=5.5cm]{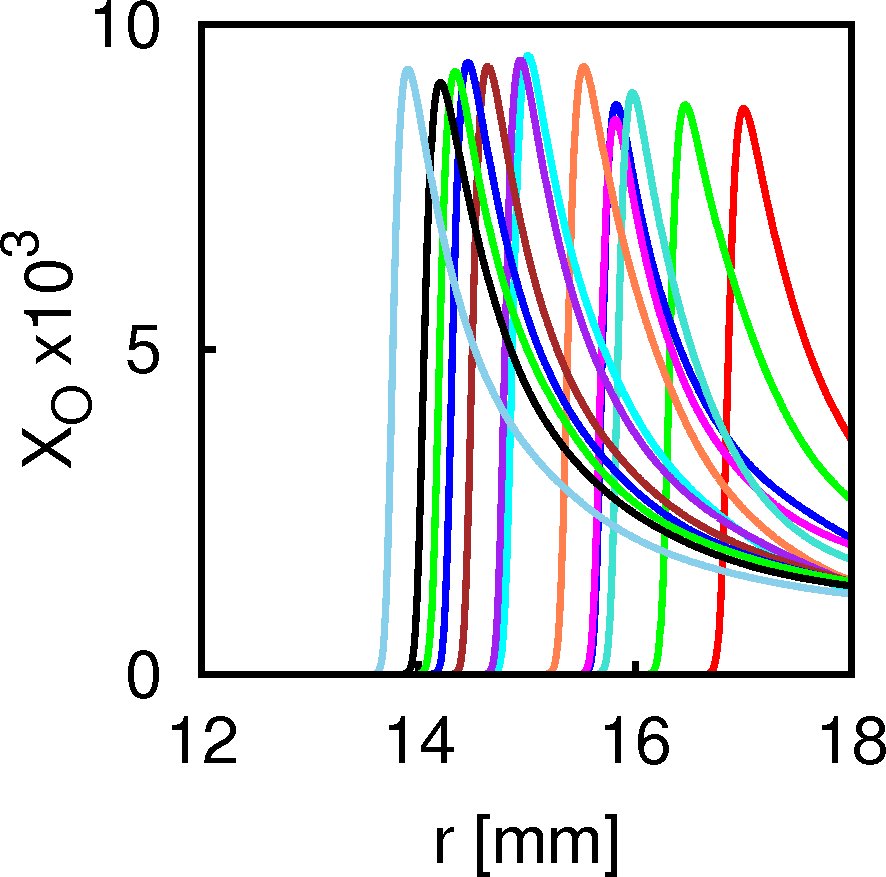}} }
\caption{As Figure 8, but for a spherical flame.}
\label{fig:F9}
\end{figure}

In \cite{MALIK2010, MALIK2012a}, some evidence of time--lag between 
the  rate of heat release and the species  response was observed in 
the flames with zero and positive stretch. A correlated response is 
more difficult to observe in the flames  with negative curvature 
in figures~8 and 9 because the flames are convected back and forth 
rapidly --smoothen
`flapping'  -- by  the pressure  induced velocity  fluctuations $u_p'$
(section 4.2).  The general expression for the flame front velocity is
   \begin{eqnarray} u_f &=& s_n+u_g+u'_p, \end{eqnarray} 
where $u_g$ is the gas velocity, and $\langle u_p' \rangle = 0$, where
the angled brackets represents the ensemble average.

Flapping  of  the  flame  front  by pressure induced fluctuations is 
always present, but it has rarely been observed because in expanding
flames, the most commonly studied case, the gas velocity is much 
bigger  than the fluctuating component, $|u_g| \gg |u'|$. But in the 
current contracting flames where $u_g=0$ and we have 
$u_f = s_n + u'_p$, and the rms $u'_p$ is comparable to the burning 
velocity, $|u'_p|_{\rm rms}\approx |s_n|$,  and therefore  the 
flapping  of the flame front is  much more apparent causing the 
flame fronts at different times to appear scrambled.

Nevertheless, an average trend is  observable in Figures 8 and 9,
with the $H$ and $O$ radical concentrations approximately  
anti-correlated (most likely  due to the main  chain branching  
reaction $H+O_2=O+OH$),  and the  peaks  in the respective  profiles   correlating  quite  well   indicating  a  short time--lag.

The results  from the previous  sections have indicated  that inwardly
propagating  flames interact  weakly with  pressure  oscillations, and
therefore we can expect that these flames will behave more like planar
flames, and therefore the chemical balance inside the
flame  kernel will remain well correlated  to the  rate  of heat
release with short time-lags.   The average trend noted above provides
some support for this.

\section{Discussion}

In this study the  transient response of premixed negatively stretched
hydrogen/air  flames  at  mean   atmospheric  pressure  and  at  mean
stoichiometry, subjected simultaneously to equivalence ratio 
variations and to naturally 
decaying pressure fluctuations at frequencies and  length scales 
that can couple to the flame  structure has been  explored numerically 
using  an implicit method well suited to resolve stiff and unsteady 
chemically reacting systems.

A    comparison   with   the    positively   stretched    H2/air 
\cite{MALIK2010} and CH4/air flames  \cite{MALIK2012a}  was
carried  out,  and  a  number  of  important features have emerged.    
Pressure  oscillations  in  the  range $200-1000$Hz produce steeper 
spectral broadening which scale close to $E_p\sim\omega^{-3}$, 
as compared to the corresponding  positively  stretched H2/air 
flame spectra which scaled close to $E_p\sim\omega^{-2}$. This   
indicates a  weaker flame-pressure response in 
negatively stretched flames, although  the   spectral  broadening   
occurs across a wider range of frequencies; in the case of imploding
spherical  flames the range of excited frequencies extends to very 
high  frequencies. It is interesting that the expanding CH4/air 
flames also produced spectra close to $E_p\sim\omega^{-3}$.

The non-dimensional relaxation number $n_R$ is defined as the ratio 
of the relaxation time to a flame time scale $ n_R={\tau_R/\tau_L}$; 
it is  a function of stretch, molecular transport processes, fuel 
distribution, and the frequency and amplitude of the
fluctuations.   $n_R$ contains  information about  the
competing processes of flame propagation and molecular transport: 
if molecular diffusion is dominant then we  would expect $n_R$ to be
small. $n_R$ may also be indicative of flame stability to external
disturbance in the sense that a  smaller  values  of  $n_R$  means
that the reactive-diffusive system returns to mean conditions 
quickly indicating  greater  flame stability.

The relaxation  number $n_R$ was found  to decrease by  only 10\% with
increasing negative stretch (flow divergence) in H2/air flames 
compared  to the  zero stretch planar flame case; but $n_R$ was found 
to decrease  by about 40\%  with increasing  positive stretch in H2/air 
flames in \cite{MALIK2010}. Negative stretch thus appears to couple 
less strongly with dynamic processes such as pressure fluctuations 
and  transport processes than does positive stretch, because it
reduces the relaxation number to a lesser extent than the latter. 
This picture is consistent with the spectral analysis mentioned above.

This point is further strengthened by the fact that the trends in 
$n_R$ in both the H2/air and CH4/air {\em expanding} flames is similar, 
see Table~2, even though H2/air and  CH4/air are at opposite ends of the 
'reactivity' scale, H2/air being a very reactive mixture.
$n_R$ thus appears to more strongly correlated with the sign of the
curvature than with the thermo-chemistry of different flame types.
This may in part be due to the fact 
that both curvature and strain contribute to positive stretch in
expanding  flames,  but  curvature  alone contributes  to
negative  stretch in imploding flames  because of the absence of strain
(gas  velocity) ahead of the flame, $u_g=0$.


$n_R$  is potentially important as a non-dimensional time  for an
inhomogeneous system to return to the mean conditions when subjected to
external  perturbations.   It  may  have significance  for  automotive
engines where such  conditions are prevalent; if a  system has a short
relaxation  time  then  it  attains  to mean  conditions  rapidly  and
therefore mean field modeling strategies could be effective.

For  moderate levels of inhomogeneity  $A_\phi\ll 1$, and short length
scales $\lambda=1mm$ and pressure  fluctuations in the frequency range
$\omega\sim  200-1000Hz$,   the  molecular  transport   processes  are
competitive  with  flame propagation  and  partially  reduce the  fuel
inhomogeneities during the time that  the flame moves through the fuel
oscillation cycle, such  that the upper and lower  limits of the flame
speed  $s_n$ are  not  attained.  Negative  curvature  appears not  to
alter   this  process  significantly under  the   conditions  of  the
simulation, which again is consistent with the spectral
and the relaxation number analysis, which both indicate 
reduced strengths of flame-pressure coupling with negative 
curvature.

Flapping of the flame induced by the pressure fluctuations is much
more  prominent in  inwardly  propagating  flames due to the zero gas
velocity  ahead of  the flame  $u_g=0$. This  is  seen  in the
scrambling of the flame fronts in successive short time frames seen in
Figures~8 and 9. Flapping has rarely been observed before because
the large gas velocity in expanding flames, the  most commonly
observed case, overwhelms the flapping since $|u_g| \gg |u'|$.

Flapping  may have  consequences for  other effects  such  as spectra,
relaxation  numbers and  time-lag through  non-linear coupling  of the
flame-front  with  the   chemical  kinetics  and  molecular  transport
processes.   But in  view  of the  weaker  flame-pressure coupling 
observed  in negatively stretched flames  it is reasonable to assume that
the  effect on time-lag will also be weaker so that  the chemical  
balance in  the flame kernel will remain close to the zero stretch 
(planar flame) case. The flame structure will then be
well  correlated  with the  rate  of heat  release and this will lead 
to a  short time-lag  as  the flame  attempts  to  adjust  to the  
unsteady  local conditions. Such a trend in the mean is observed over 
short times in Figures~8 and 9.

However, it is possible that  different conditions -- such as stronger
pressure  fluctuations,  stronger  inhomogeneity,  stronger stretch,
different fuels,  and lean/rich conditions  -- could produce  stronger 
or  weaker  disruption to  the
chemical balance through  coupling with non-linear transport processes
and  pressure  fluctuations. To  some  extent  this  has already  been
observed in outwardly propagating mode in \cite{MALIK2010, MALIK2012a} 
where the impact of positive stretch was stronger.
\cite{JOHAN2011} have looked at turbulent premixed hydrogen flames and
observed dominant kinetics or dominant pressure effects,
depending upon fuel-lean and fuel-rich conditions. However, it is not
yet known whether this is specific to hydrogen flames.

The  response  of the  thermochemical  structure  to pressure and
fuel distribution fluctuations is  similar to  that
observed in the expanding flames in \cite{MALIK2010, MALIK2012a}. The 
evidence suggests that we can  characterise flame  response by a 
spectrum of  non-dimensional length scales  $\{S_k=l_k/\lambda\} 
\quad k=1,2,..., N$, where $l_k$  is the lengths scale of the 
$k'th$ species, and  $\lambda$  is the  scale of
the  physical oscillations (pressure  oscillations  in  the  current
study).  We  find  that   all  species   oscillate  in   their  peak
concentrations  in  response  to  the local  pressure  and  fuel
distribution; but thick unstable  species with $S_k\gg 1$ are also 
disturbed across their  entire length.  The major  stable species 
concentrations are `saturated'  at high levels  and the induced 
fluctuations  are too small to be observable in their profiles.

The study reported here is of importance to  combustion processes at a
fundamental level as  well as to practical devices  such as automotive
engines and gas turbines.   Engines for example increasingly use
lean fuel mixtures and there is also a trend towards the use of
alternative fuels, such as biofuels. This requires the  study of 
lean combustion  processes for the  prediction  of  emissions  levels,  
which in  turn  requires  the inclusion of realistic chemistry 
coupled to the compressible flow. The method used here demonstrates 
that detailed numerical studies of such complexity is now possible.

\section{Conclusions}

The implicit method TARDIS, \cite{MALIK2010,MALIK2012a, MALIK2012b}, 
has provided a  platform to  explore finite thickness premixed  flames  
in great  detail because of its capability of coupling the fully
compressible  flow   to  the  comprehensive (realistic)  chemistry  
and detailed  transport properties. The implicit nature of the solver
gives  it stability,  and it  resolves  all the  temporal and spatial
scales  in the  systems examined here. This feature allows very complex
systems in unsteady contexts to be explored, and as such is a step 
towards more realistic combustion simulations.

We have explored negatively stretched hydrogen/air  flames subjected
to  fluctuations in pressure  and  fuel  distributions. The impact of
increasing  negative  stretch  was  investigated through  the  use  of
planar, cylindrical  and spherical  geometries, and a  comparison with
the results from expanding hydrogen/air flames in 
\cite{MALIK2010,MALIK2012a} was made. Imploding flames are of interest  
because they have rarely been studied and also because only
negative curvature  contributes to the stretch so that the effects of
curvature can be studied in isolation.

Overall, we have found that the results from axi-symmetric (cylindrical) 
flames are qualitaively similar to those from the corresponding spherical 
flames. This is consistent with the previous studies for expanding
H2/air flames, \cite{MALIK2010}, and expanding methane/air flames,
\cite{MALIK2012b}. This suggests that it is the fuel type and the flame propagation mode that are the dominant features in the these systems.

The  flame relaxation number
$n_R=\tau_R/\tau_L$, (where  $\tau_R$ is the time that  the flame speed
$s_n$ takes to return to the mean equilibrium conditions after initial
disturbance; $\tau_L$  is a flame time scale)  contains information
about   competing  processes  of   flame  propagation   and  molecular
transport. $n_R$  decreases by  only 10\%  with increasing
{\em negative} stretch, and by about 40\% in expanding flames with
{\em positive} stretch in~\cite{MALIK2010, MALIK2012a}, see Table~2.

Furthermore,  the  fluctuating  pressure spectra  $E_p(\omega)$  scale
close  to $\sim\omega^{-3}$, which  is steeper  than in  the expanding
H2/air flames  which  scales  like  $\sim\omega^{-2}$, but is close to
the spectra observed in the expanding CH4/air flames.

The flame relaxation number $n_R$ appears to be far more sensitive to 
variations in positive and negative curvature than it is to the different
thermo-chemistry in different flame types. The decrease in $n_R$ is much 
less with negative curvature than with positive curvature. $n_R$ may 
therefore be an indicator of flame-curvature coupling, and the results
here may indicate that negative curvature couples less strongly to the 
flame thermo-chemistry than in positive curvature flames, which is 
consistent with the spectral analysis mentioned
above. However this needs more investigation and is the subject of 
current ongoing work.

Rapid  flapping of  the flame  front  by the  random velocity fluctuations 
induced by the pressure fluctuation is prominent  again due to the lack 
of  gas velocity 
ahead of the flame. `Memory effects' between fuel  consumption and the
rate  of  heat  release  is  obscured by  the  flapping,  although  on
theoretical  grounds we  expect  inwardly propagating  flames to  have
shorter time-lags than in positively stretched flames.

A  spectrum of species  thickness $\{S_k, k=1,2..,N\}$  exists such  that species
formed in thin reaction layers are generally not disturbed except in
their peak concentrations. Stable species profiles are unaffected by
small levels of oscillations. Thus,  it may be
more useful to characterise the internal structure of flames in terms 
of a spectrum of length scales rather than in terms of a single 
flame thickness.

\subsection*{Acknowledgment}

The authors wish to thank SABIC for funding this work through
grant number SB101018, and also the Information Technology Center at 
King Fahd University of Petroleum and Minerals for providing High 
Performance Computing resources that have contributed to the 
results reported in this paper.

\nolinenumbers

\end{document}